# Coalescing neutron stars – a step towards physical models

## I. Hydrodynamic evolution and gravitational-wave emission

**M. Ruffert[1]★, H.-Th. Janka[1,2]★★, and G. Schäfer[3]★★★**

[1] Max-Planck-Institut für Astrophysik, Karl-Schwarzschild-Str. 1, Postfach 1523, 85740 Garching, Germany
[2] Department of Astronomy and Astrophysics, University of Chicago, 5640 S. Ellis Avenue, Chicago, IL 60637, USA
[3] Max-Planck-Arbeitsgruppe "Gravitationstheorie", Friedrich–Schiller–Universität, Max-Wien-Platz 1, 07743 Jena, Germany

September 1, 1995

**Abstract.** We investigate the dynamics and evolution of coalescing neutron stars. The three-dimensional Newtonian equations of hydrodynamics are integrated by the "Piecewise Parabolic Method" on an equidistant Cartesian grid with a resolution of $64^3$ or $128^3$. Although the code is purely Newtonian, we do include the emission of gravitational waves and their backreaction on the hydrodynamic flow. The properties of neutron star matter are described by the physical equation of state of Lattimer & Swesty (1991). In addition to the fundamental hydrodynamic quantities, density, momentum, and energy, we follow the time evolution of the electron density in the stellar gas. Energy loss by all types of neutrinos and changes of the electron fraction due to the emission of electron neutrinos and antineutrinos are taken into account by an elaborate "neutrino leakage scheme". We simulate the coalescence of two identical, cool neutron stars with a baryonic mass of $\approx 1.6\,M_\odot$ and a radius of $\approx 15$ km and with an initial center-to-center distance of 42 km. The initial distributions of density and electron concentration are given from a model of a cold neutron star in hydrostatic equilibrium, the temperature in our initial models is increased such that the thermal energy is about 3% of the degeneracy energy for given density and electron fraction (central temperature about 8 MeV). We investigate three cases which differ by the initial velocity distribution in the neutron stars, representing different cases of the neutron star spins relative to the direction of the orbital angular momentum vector. The orbit decays due to gravitational-wave emission, and after half a revolution the stars are so close that dynamical instability sets in. Within about 1 ms they merge into a rapidly spinning ($P_{\rm spin} \approx 1$ ms), high-density body ($\rho \approx 10^{14}$ g/cm$^3$) with a surrounding thick disk of material with densities $\rho \approx 10^{10} - 10^{12}$ g/cm$^3$ and orbital velocities of 0.3–0.5 c. In this work we evaluate the models in detail with respect to the gravitational wave emission using the quadrupole approximation. In a forthcoming paper we will concentrate on the neutrino emission and implications for gamma-ray bursters. The peak emission of gravitational waves is short but powerful. A maximum luminosity in excess of $10^{55}$ erg/s is reached for about 1 ms. The amplitudes of the gravitational waves are close to $3 \cdot 10^{-23}$ at a distance of 1 Gpc, and the typical frequencies are between 1 KHz and 2 KHz, near the dynamical frequency of the orbital motion of the merging and coalescing neutron stars. In contrast to the diverging gravitational wave amplitude of two coalescing point-masses, our models show decreasing amplitudes of the waves because of the finite extension of the neutron stars and the nearly spherical shape of the merged object toward the end of the simulations. The structure and temporal development of the gravitational wave signal and energy spectrum show systematic trends with the amount of angular momentum in the system and depend on the details of the hydrodynamic mass motions.

## 1. Introduction

Gravitational wave astronomy may soon become an observational science, since three gravitational wave experiments, LIGO (Abramovici et al, 1992), VIRGO (Bradaschia et al, 1991), and GEO600 (Danzmann et al, 1995a) with four interferometric detectors are under construction. Moreover, the European Space Agency (ESA) has chosen the space-based LISA laser interferometer (Danzmann et al, 1995b) to be a cornerstone project. All sources are expected to be astrophysical, with merging and coalescence of neutron stars and black holes as well as rapidly rotating neutron stars and supernovae being the most promising candidates (Thorne, 1992). Whether these events and objects will be detectable or not does not only depend

---

on their occurrence rates and the strength of the emitted waves, but it also depends on the frequency of the gravitational waves relative to the resonance frequencies of the detectors (e.g. $f \approx$ 10–1000 Hz for terrestrial detectors) (Finn & Chernoff, 1993; Finn, 1994). Only an a priori knowledge of the structure of typical gravitational wave signals will allow one to extract the faint signals from a very noisy background. The event rates are highly uncertain and depend on the Hubble constant, pulsar and supernova rates, and orbital parameters of binary systems. Very optimistic estimates yield several neutron star and black hole mergers per year within about 25 Mpc, while pessimistic numbers quote a few per year within 1 Gpc (e.g. Clark et al, 1979; Phinney, 1991).

Two directions of theoretical investigations have been followed up. (a) Calculations of the orbital decay of binaries due to gravitational radiation, treating the binary components as point-masses and using progressively more elaborate post-Newtonian approximations to the equations of general relativity. These computations can integrate the orbit over lots of revolutions. Examples of such investigations are, e.g., Lincoln & Will (1990), Iyer & Will, (1993), Cutler & Flanagan, (1994), Imshennik & Popov (1994). (b) Hydrodynamic simulations that concentrate on the merging phase of the binary and the last few orbital revolutions before the merging. These simulations take into account the finite extension of the binary components and allow one to study the fluid dynamics effects and to include detailed microphysics. In particular, computations with different numerical methods, with different handling of the gravitational wave emission and of the backreaction on the hydrodynamic flow, and with different parameters in a polytropic equation of state have been performed so far (e.g., Shibata, Nakamura & Oohara, 1993; Zhuge, Centrella & McMillan 1994; Rasio & Shapiro 1994; Davies et al, 1994; also references cited in these papers). Our work intends to go a step further in the numerical modeling of the hydrodynamic phase of the inspiral by using a complex, "physical" equation of state (Lattimer & Swesty, 1991) and by including the effects of local composition changes and cooling due to neutrino emission.

Gravitational waveforms and luminosities were calculated in the post-Newtonian expansion for the small parameter $\epsilon \equiv (v/c)^2$ to terms of order $\mathcal{O}(\epsilon^{5/2})$ by Lincoln & Will (1990). Iyer & Will (1993) derived the post-Newtonian radiation reaction terms at $\mathcal{O}(\epsilon^{7/2})$. Imshennik & Popov (1994) showed that independent of the initial parameters of the binary orbit the final orbital eccentricity gets very small. Cutler & Flanagan (1994) investigated which, and how accurately, binary parameters can be deduced from the waveforms. They found, e.g., that although to lowest post-Newtonian order the phase depends only on the "chirp mass" $\mathcal{M} = (M_1 M_2)^{3/5}(M_1 + M_2)^{-1/5}$ which is measurable to a relative accuracy better than 1%, higher terms also involve the reduced mass.

In a sequence of papers a Japanese group (Shibata, Nakamura & Oohara, e.g., 1993) reported about hydrodynamic simulations with different initial conditions of the neutron stars' spins, masses, and tidal deformations and with different shapes of the orbit. The maximum amplitude of the gravitational radia-tion at 10 Mpc turned out to be $1$–$5 \cdot 10^{-21}$ and the total amount of energy emitted in gravitational waves was determined to be 1–3% of the total rest mass. Especially the latter values are strongly dependent on the size and compactness of the neutron stars since the energy in gravitational waves increases like $1/r^5$ with the minimum separation of the neutron stars before they merge and the binary system loses its highly non-spherical geometry.

Zhuge et al (1994) focussed on the energy spectrum that gives the energy emitted in gravitational waves at different frequencies. They discussed the information about the dynamics of the coalescence and about the merging objects that may be provided by the spectrum. Certain stages of the binary evolution are reflected in particular wave features. From their analysis it is evident that the equation of state and the neutron star masses and radii influence the frequencies and amplitudes of spectral structures. Therefore an interpretation of details and fine structure in gravitational wave signals seems to require models of the merger scenario that include a physically more meaningful description of neutron star matter than by the usually employed polytropic equations of state. Microphysical processes at high densities, like changes of the $\beta$-equilibrium, give rise to bulk viscosity and neutrino emission and must be suspected to cause dissipation and to smooth out or modify some of the fine structure found in idealized models. The same effect may be associated with the gravitational radiation (back)reaction which was also not taken into account by Zhuge et al (1994).

Rasio & Shapiro (1994) found that triaxial configurations are formed by the merging of neutron stars with a sufficiently stiff equation of state. Non-axisymmetric structures occur for adiabatic exponents of 3 or higher (polytropic index 0.5 or below) in the polytropic equation of state. In this case the peak amplitude of the gravitational wave emission is substantially larger and the emission proceeds for a longer time after the coalescence. Using hydrodynamic models that were computed with a polytropic equation of state Davies et al (1994) tried to estimate the temperatures in the central, merged object and in the surrounding thick disk and obtained values of around 10 MeV and 3 MeV, respectively. Based on these estimates they attempted to discuss possible implications for neutrino emission, gamma-ray bursts, and thermonuclear reactions and energy generation in the disk.

The hydrodynamic simulations presented in this work employ the high-density equation of state of Lattimer & Swesty (1991) which describes the thermodynamics of the neutron star matter in dependence of density, temperature, and composition, the latter being expressed by the electron fraction $Y_e$. Changes of $Y_e$ and cooling of the matter due to the emission of neutrinos of all types are taken into account. The use of a microphysically meaningful equation of state and the consideration of neutrino source terms imply that dissipation effects due to the bulk viscosity of the medium are included in our models. Although our computations, like those of the hydrodynamic approaches summarized above, were performed with a basically Newtonian code, we add the terms into the hydrodynamic equations that describe the effects of gravitational wave emission and the corresponding backreaction on the hydrodynamic flow. This is done according to the formalism of Blanchet et al (1990) by following the special implementation chosen by Shibata et al (1992). Certainly, not taking into account the gravitational radiation reaction (Rasio & Shapiro, 1994) is a poor choice. Our treatment is also more accurate than the point-mass quadrupole approximation that was used by Davies et al (1994) and Zhuge et al (1994) for the phase of the coalescence when the neutron stars were still separated and switched off when the merging happened.

Our simulations are performed with an explicit, Eulerian finite-difference scheme based on the PPM-method of Colella & Woodward (1984). This grid-based scheme is superior to a particle-based method like SPH in the handling of shock waves, which is done very well by solving local Riemann problems. Compared with other Eulerian algorithms the PPM method has also a rather small numerical viscosity. This seems to be demanded since recent work (Kochanek, 1992; Lai, 1994; Reisenegger & Goldreich, 1994; see Sect. 3 for a discussion) suggests that the viscosity of neutron star matter is so small that the stars cannot develop synchronous rotation during inspiral.

Our models yield detailed information about the neutrino emission (lepton number as well as energy) from the merging stars. This allows us to analyse the merger scenario not only as a source of gravitational radiation but also to perform a quantitative investigation of the possibility to power gamma-ray burst events by the energy deposition due to neutrino-antineutrino annihilation in the surroundings of the merging binary. We shall report results of this part of our work in a subsequent paper, while we will concentrate on the gravitational wave aspect here. Section 2 provides a description of the numerical methods and of the treatment of the microphysics. In Sect. 3 we explain the set of computed models and give details about the initial conditions our simulations are started from. The results are presented in Sect. 4, followed by a discussion in Sect. 5.

## 2. Computational procedure

### 2.1. Hydrodynamics, self-gravity and gravitational-wave backreaction

The scenario of coalescing neutron stars is a three-dimensional problem with the orbital plane being a plane of symmetry. We perform the simulations on an equidistant Cartesian grid with the number of zones being 64 or 128 in both space dimensions of the orbital plane. Perpendicular to the orbital plane (i.e. in the z-direction) we use only 1/4 of this number of zones. One factor of two is saved due to the symmetry about the orbital plane: we only simulate the volume on one side of this plane. A second factor of two can be saved by only using a region vertical to the orbital plane which has a quarter of the size of the region modeled in the orbital plane. This is possible because test calculations showed that hardly any matter moves out to more than one neutron star radius away from the orbital plane.

The neutron star matter is evolved hydrodynamicly via the Piecewise Parabolic Method (PPM) developed by Colella and Woodward (1984). Details of the actual implementation of the PPM code can be found in Fryxell et al (1989) and in Ruffert (1992). It is Eulerian and explicit in time and includes the non-constant-$\gamma$ equation of state formalism as described by Colella & Glaz (1985). The code is basically Newtonian, but was extended to contain the terms necessary to describe gravitational waves and their backreaction (Blanchet et al, 1990). We use a form similar to Shibata et al (1992) in notation and in use of the energy equation instead of the enthalpy equation. However, we employ the total specific energy, i.e. the sum of specific internal energy and specific kinetic energy as fundamental hydrodynamic variable. The hydrodynamic laws of conservation of mass, momentum and energy for non-viscous flow, in ADM coordinates (generalized isotropic coordinates), are, respectively,

$$\frac{\partial \rho}{\partial t} + \frac{\partial \rho v^j}{\partial x^j} = 0 \quad , \tag{1}$$

$$\frac{\partial \rho w^i}{\partial t} + \frac{\partial (\rho w^i v^j + P \delta^{ij})}{\partial x^j} = -\rho \frac{\partial \psi}{\partial x^i} - \rho \frac{\partial \phi}{\partial x^i} \quad , \tag{2}$$

$$\frac{\partial \rho E}{\partial t} + \frac{\partial (\rho E + P) v^j}{\partial x^j} = -\rho v^j \frac{\partial \psi}{\partial x^j} + W + S_{\rm E} \quad , \tag{3}$$

with the energy source term due to gravitational waves given by

$$W = -\rho v^i \frac{\partial \phi}{\partial x^i} + \frac{4}{5} \frac{G}{c^5} \dddot{D}_{ij} \, v^i \left( \rho \frac{\partial \psi}{\partial x^j} + \frac{\partial P}{\partial x^j} \right) \quad . \tag{4}$$

The symbols have the usual meanings: $\rho$ denotes mass density, $t$ time, $v^i$ and $w^i$ the components of the kinematic and dynamic velocity, respectively, $x^i$ position vector components, $P$ pressure, $\psi$ the Newtonian gravitational potential, $\phi$ the backreaction potential due to gravitational waves, $E$ is the total specific energy (sum of specific internal energy $\epsilon$ and specific kinetic energy $\frac{1}{2} w^i w^i$), $G$ and $c$ the gravitational constant and the speed of light, $\dddot{D}_{ij}$ the third time derivative of the quadrupole moment, and $S_{\rm E}$ the energy loss due to neutrino emission. Details of $S_{\rm E}$ follow in Sect. 2.4 and Appendix B. The total emitted luminosity of gravitational waves $\mathcal{L}$ can be obtained either by summing up $W$ over the whole emitting volume or via the classical quadrupole formula

$$\mathcal{L} = \frac{1}{5} \frac{G}{c^5} \dddot{D}_{ij} \dddot{D}_{ij} \quad . \tag{5}$$

Notice that summing up $W$ yields the gravitational wave luminosity without averaging over time. When the orbit decays and therefore is not perfectly circular, the non-averaged luminosity is in general not identical with the value obtained by averaging over one orbital period. All secondary quantities are calculated from the primary quantities $\rho$, $\rho w^i$, and $\rho E$ (used in the code) by the following relations

$$E = \epsilon + \frac{1}{2} w^i w^i \quad , \tag{6}$$

$$v^i = w^i + \frac{4}{5}\frac{G}{c^5}\ \dddot{D}_{ij}\ w^j \quad , \tag{7}$$

$$\phi = \frac{2}{5}\frac{G}{c^5}\left(R - \dddot{D}_{ij}\ x^j\frac{\partial\psi}{\partial x^i}\right) \quad , \tag{8}$$

$$\dddot{D}_{ij} = \text{STF}\left[2\int dV\right.$$
$$\left.\left(2P\frac{\partial v_i}{\partial x^j} + \frac{\partial\psi}{\partial x^j}(x_i\frac{\partial\rho v^k}{\partial x^k} - 2\rho v_i) - \rho x_i\frac{\partial\dot\psi}{\partial x^j}\right)\right] \quad . \tag{9}$$

The notation STF means *symmetric* and *trace free*,

$$\text{STF}\left[X_{ij}\right] \equiv \frac{1}{2}X_{ij} + \frac{1}{2}X_{ji} - \frac{1}{3}\delta_{ij}X_{kk} \quad , \tag{10}$$

and the following three Poisson equations have to be solved:

$$\Delta\psi = 4\pi G\rho \quad , \tag{11}$$

$$\Delta\dot\psi = -4\pi G\frac{\partial\rho v^i}{\partial x^i} \quad , \tag{12}$$

$$\Delta R = 4\pi G\ \dddot{D}_{ij}\ x^j\frac{\partial\rho}{\partial x^i} \quad . \tag{13}$$

The Poisson equations in integral form are interpreted as convolution and calculated by fast Fourier transform routines: non-periodic grid boundaries are enforced by zero-padding (e.g. Press et al, 1992; Eastwood & Brownrigg, 1979). This doubles the number of Fourier components compared to the original number of grid zones per dimension. The accelerations that follow from the potential are added as source terms in the PPM algorithm. All spatial derivatives necessary for the gravitational-wave terms are implemented as standard centered differences on the grid.

We implement the hydrodynamic laws including gravitational backreaction (Eqs. 1 to 3) in the following way. First the PPM routines produce the kinematic velocities at the zone interfaces by solving Newtonian Riemann problems. Then these kinematic velocities are transformed to dynamic velocities (Eq. 7) and the relativistic momentum ($\rho w^i$) and total specific energy $E$ are advected. Finally the new relativistic quantities are transformed back to kinematic velocities to be used as start values for the next explicit PPM step.

## 2.2. Equation of state

In general, the state of matter in equilibrium is characterized by three independent quantities, e.g. baryon (rest mass) density $\rho$, electron fraction $Y_e$, and internal energy density $e = \epsilon\rho$. All other thermodynamic quantities, e.g. pressure $P$, temperature $T$, etc., can be computed from these fundamental quantities. Since we include neutrino processes, the electron number density $n_e$ has to be advected, too:

$$\frac{\partial n_e}{\partial t} + \frac{\partial n_e v^j}{\partial x^j} = S_L \quad . \tag{14}$$

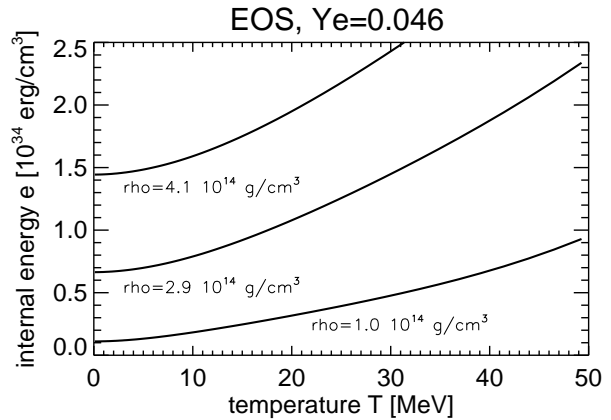

**Fig. 1.** An excerpt of the equation of state: the internal energy density as a function of temperature for three densities at one specific value of $Y_e$.

The electron number fraction $Y_e$ can be recovered from

$$Y_e = \frac{n_e}{\rho}u \quad , \tag{15}$$

where $u$ denotes the atomic mass unit. Information about the lepton source term $S_L$ can be found in Sect. 2.4 and Appendix B.

We use the equation of state of Lattimer & Swesty (1991) in tabular form (incompressibility parameter of bulk nuclear matter $K$=180 MeV). The tabulated quantities are the pressure $P$, the internal energy density $e$, the adiabatic index $\Gamma = (\partial\ln P/\partial\ln\rho)|_s$, and the degeneracy parameters of protons and of neutrons (without rest masses). The tables span the following ranges: (a) density: $9 \lesssim \lg\rho$ [g/cm$^3$] $\lesssim 15.5$ in 130 steps, (b) the temperature: $-1.0 \lesssim \lg T$ [MeV] $\lesssim 1.7$ in 108 steps, and (c) the electron number fraction: $0.006 \lesssim Y_e \lesssim 0.49$ in 25 steps. First a large table was generated with double as many entries in every dimension. The entries at small values of $Y_e$ were obtained from extrapolation, since the original Lattimer & Swesty (1991) FORTRAN version is not applicable for $Y_e \lesssim 0.03$. Although Lattimer & Swesty (1991) modeled the equation of state to arbitrarily high densities, they did not consider possible additional physics such as pion or kaon condensation, hyperons, and the quark-hadron phase transition. Therefore their equation of state provides a physical description of nuclear matter only below about twice nuclear density. Note, however, that only relatively small regions in our $M \approx 1.6\,M_\odot$ neutron star models and in our simulated mergers actually reach densities above this value. If the FORTRAN version of the equation of state of Lattimer & Swesty (1991) did not yield converged values, the tabulated quantities were interpolated from neighbouring entries. Non-monotonic values of the energy dependence on temperature were smoothed by averaging over neighbouring entries in the table. Unambiguous inversion of the equation of state requires strictly monotonic behaviour. Also isolated positive and negative spikes of the adiabatic index $\Gamma$ were reduced: negative ones by replacing the entries with $\Gamma = 0.01$ and spikes over $\Gamma = 4$ were replaced by the average of six neighbouring entries in the table. Finally, the reduced table version (with the binning (a), (b), and (c) as specified above) was obtained by choosing new grid points in the lg $\rho$-lg $T$-$Y_e$-space which are cubic-centered between the points of the large table and attributing to them the averages of the eight ($=2^3$) closest neighbours of the large table. This reduces the amount of data and smooths the functions further.

The basic quantities used by the PPM code are density $\rho$, electron concentration $Y_e$, and energy $E$. Thus the tables have to be inverted to obtain the temperature. This is done by a bisection iteration. The interpolations in the tables are performed trilinearly in lg $\rho$, lg $T$ and $Y_e$. A particular problem with the equation of state arises in the very degenerate, low-temperature regime, where the internal energy density $e$ varies only weakly with temperature. Therefore small variations of $e$ imply large shifts of $T$. Exemplary curves of $e(T)$ are plotted in Fig. 1.

The inversion of the table for typical values of density, energy and $Y_e$ was checked by first computing the corresponding temperatures and then using the table a second time in reverse direction to regain the original quantities. The relative difference between these values and the original ones amounts to approximately machine accuracy ($10^{-12}$) in all tested cases. Because of the numerical problems and the indispensable smoothing steps described above we consider the employed tabular resolution to be sufficient and to ensure the thermodynamical consistency of the equation of state to an acceptable degree.

### 2.3. Gravitational wave emission

Apart from the gravitational wave luminosity (Eq. 5), the amplitudes and waveforms of the gravitational wave emission are of immediate interest, because they are the quantities that determine the measurements. The amplitudes of the two physical (i.e. coordinate–invariant or gauge–invariant) polarizations of gravitational waves that are emitted perpendicular to the orbital plane and are observed at a distance $r$ are given by

$$h_+ = \frac{G}{c^4} \frac{1}{r} \left( \ddot{D}_{xx} - \ddot{D}_{yy} \right) \tag{16}$$

and

$$h_\times = \frac{G}{c^4} \frac{2}{r} \ddot{D}_{xy} \quad . \tag{17}$$

In the direction along the rotation axis the gravitational-wave signal is largest. The amplitude $h_+$ of the radiation in direction $(\theta, \varphi)$ then follows to (Rasio & Shapiro, 1994)

$$h_+(\theta, \varphi) = -h_+(0,0) \cos\theta \sin 2\varphi + h_\times(0,0) \cos\theta \cos 2\varphi , \tag{18}$$

where $h_+(0,0)$ and $h_\times(0,0)$ are the amplitudes along the rotation axes, as given by Eqs. 16 and 17, respectively. $h_\times(\theta, \varphi)$ cannot be simply written as function of $h_+(0,0)$ and $h_\times(0,0)$ only.

We calculate the second time derivative of the quadrupole moment from

$$\ddot{D}_{ij} = \text{STF} \left[ 2 \int dV \rho \left( v_i v_j - x_i \frac{\partial \psi}{\partial x_j} \right) \right] \quad . \tag{19}$$

Neither in these equations, nor in those pertaining to the source terms necessary in the hydrodynamic code (Eqs. 4, 9, etc.), need time derivatives be calculated explicitly. Therefore, smoothing of gravitational wave quantities is unnecessary and all plots containing gravitational wave properties are done without any smoothing, except where noted.

Analytic expressions for quasi–circular inspiral can be derived in the point–mass approximation (e.g. Misner et al, 1973). The separation $a(t)$ of the two point masses, each with mass $M$, as a function of time $t$ is

$$a(t) = a_0 \left( 1 - \frac{t}{t_0} \right)^{1/4} \quad \text{with} \quad t_0 = \frac{5}{64} \frac{a_0^4}{R_s^3 c} \quad . \tag{20}$$

$t_0$ is a function of the separation $a_0$ at time $t = 0$. $R_s = 2GM/c^2$ is the Schwarzschild-radius of mass $M$. The gravitational waves emitted then follow to (e.g. Lai et al, 1994)

$$h_+^{\rm p}(\theta = 0, \varphi, t) = -\frac{1}{r} \frac{R_s^2}{a(t)} \sin(\Phi(t) - \varphi) \quad , \tag{21}$$

and

$$h_\times^{\rm p}(\theta = 0, \varphi, t) = -\frac{1}{r} \frac{R_s^2}{a(t)} \cos(\Phi(t) - \varphi) \quad , \tag{22}$$

where the superscript p indicates the point-mass result. The angle $\Phi(t)$ as a function of time is given by

$$\Phi(t) = \Phi_0 - \frac{1}{4} \left( \frac{a(t)}{R_s} \right)^{5/2} \quad , \tag{23}$$

where $\Phi_0 = \Phi(t = t_0)$ is the final angle between the axis connecting the two point masses and the direction $\varphi = 0$. The power radiated away in gravitational waves is (e.g. Shapiro & Teukolsky, 1983)

$$\mathcal{L}^{\rm p} = \frac{2}{5} \frac{c^5}{G} \frac{R_s^5}{a^5} \quad . \tag{24}$$

The gravitational wave luminosity of Eq. 24 represents a time-averaged value. Note that for circular orbits, the local power expression Eq. 5 yields Eq. 24 even without averaging over one orbital period. Our results concerning the gravitational wave luminosity (Fig. 22) are somewhat coordinate dependent, especially during the dynamic "plunge", although the total amount of gravitational wave energy emitted ($\mathcal{E}$ in Table 1) as well as the energy spectrum (Fig. 27) are independent of the coordinate system.

The energy emitted in gravitational waves per unit frequency interval $d\mathcal{E}/df$ is given (Zhuge et al, 1994; Thorne, 1987; Misner et al, 1973, Box 37.4.C) by

$$\frac{d\mathcal{E}}{df} = \pi^2 f^2 \frac{G}{c^5} \frac{8}{15} \left( 6|C_{xy}|^2 + 6|C_{yz}|^2 + 6|C_{xz}|^2 \right. \tag{25}$$
$$\left. + |C_{xx} - C_{yy}|^2 + |C_{yy} - C_{zz}|^2 + |C_{xx} - C_{zz}|^2 \right) \quad ,$$

where $C_{ij}$ contains the Fourier transform of $\dddot{D}_{ij}$. We numerically calculate the six power spectra $C^2$ via the discrete fast Fourier transform (FFT): $C_{ij} = \mathrm{FFT}[\dddot{D}_{ij}]$. Finally, the values resulting from Eq. 25 are smoothed before they are plotted, by averaging over 3 neighbouring array elements using a top-hat function.

For point-masses in quasi–circular motion one obtains the relation (e.g. Cutler & Flanagan, 1994)

$$\frac{d\mathcal{E}^{\mathrm{p}}}{df} = \frac{R_{\mathrm{s}}^2 c^3}{G}\left(\frac{f_0}{f}\right)^{1/3} \quad \text{with} \quad f_0 = \frac{\pi^2}{12^3}\frac{c}{R_{\mathrm{s}}} \quad . \tag{26}$$

The energies given in Eqs. 26 and 26 represent averages over one orbital revolution.

### 2.4. Neutrino treatment

In the context of the three-dimensional hydrodynamic modeling of the coalescence of binary neutron stars, the production and emission of neutrinos is treated in terms of a neutrino-leakage scheme. Since the timescale of the dynamical merging event covered by the presented simulations is of the order of 10 ms only, transport effects by the diffusion of neutrinos can be expected to be of minor importance. Moreover, neutrino momentum transfer plays a negligible dynamical role. Therefore the use of a leakage scheme that takes into account the possible equilibration of neutrinos and the opaque, hot stellar matter appears adequate to describe local neutrino sources and sinks.

Electron-type neutrinos $\nu_e$ and $\bar{\nu}_e$ are created via charged-current $\beta$-processes

$$e^- + p \longrightarrow n + \nu_e \,, \tag{27}$$

$$e^+ + n \longrightarrow p + \bar{\nu}_e \,, \tag{28}$$

by electron-positron pair-annihilation,

$$e^- + e^+ \longrightarrow \nu_i + \bar{\nu}_i \,, \tag{29}$$

and – with a dominant rate at high densities and high electron degeneracy – by plasmon decays,

$$\tilde{\gamma} \longrightarrow \nu_i + \bar{\nu}_i \,. \tag{30}$$

The emission of $\nu_e$ and $\bar{\nu}_e$ extracts lepton number and energy, while heavy-lepton neutrinos $\nu_x \equiv \nu_\mu, \bar{\nu}_\mu, \nu_\tau, \bar{\nu}_\tau$, generated via the "thermal" processes of Eq. (29) and Eq. (30), carry away energy from the stellar medium. At low optical depths the production and emission of neutrinos is computed directly from the rates of the above processes. In contrast, at high optical depths the equilibrium timescales are much shorter than the timescales of neutrino diffusion and of hydrodynamic changes. Neutrinos are therefore assumed to be present in their chemical equilibrium abundances and the loss of neutrino number and energy proceeds on the diffusion timescale. For detailed information about the neutrino production rates and the superposition of the treatments in optically thick and thin regions, see Appendix B. All rates are approximated by simple analytical formulae, which allow for a very fast numerical evaluation and whose accuracy of the order of a few 10% in describing the neutrino emission of hot neutron star matter is sufficient in the context of the presented work.

The optical depth for $\nu_e$ and $\bar{\nu}_e$ is dominated by the inverse reactions of the $\beta$-processes of Eq. (27) and Eq. (28), $\nu_e$ absorption onto neutrons,

$$n + \nu_e \longrightarrow e^- + p \,, \tag{31}$$

and $\bar{\nu}_e$ absorption onto protons,

$$p + \bar{\nu}_e \longrightarrow e^+ + n \,. \tag{32}$$

An important contribution also comes from neutral-current scattering off nucleons,

$$\nu_i + \begin{Bmatrix} n \\ p \end{Bmatrix} \longrightarrow \nu_i + \begin{Bmatrix} n \\ p \end{Bmatrix} \,, \tag{33}$$

which is by far the most important opacity source for heavy-lepton neutrinos. Since for numerical reasons the employed equation of state of Lattimer & Swesty (1991) had to be used as a table (see Sect. 2.2) which we desired to keep as small as possible, we did not tabulate detailed information about the baryonic composition of the neutron star medium. Therefore we neglected neutrino scatterings off nuclei and assumed the stellar gas to be completely dissociated into free neutrons and protons in the determination of optical depths. The associated uncertainties are of the order of a few 10% and therefore within the limits of accuracy of other aspects of the neutrino description in this work. The spectrally averaged opacities for the processes of Eq. (31)–Eq. (33) are given in Appendix A.

Once the neutrino opacities are determined one can compute optical depths and diffusion timescales, which enter the computation of the "effective" lepton number and energy loss rates of the stellar gas. The latter are defined as a continuous superposition of the rates of the processes Eq. (27)–Eq. (30) at low optical depths and of a diffusion loss term that represents the neutrino leakage from optically thick regions on the neutrino diffusion timescale (see Appendix B). Adding up the neutrino emission from all zones of the grid yields total neutrino number and energy fluxes. This neutrino leakage treatment was compared with the results of neutrino diffusion calculations for cooling (spherically symmetrical) protoneutron stars and the agreement was found to be on the level of a few 10% again.

## 3. Initial conditions

The initial situation is defined by two identical neutron stars of about $M = 1.63\,M_\odot$ and 15 km radius that are placed at a center-to-center distance of $a_0 = 42$ km on a grid of size 82 km. The distributions of density $\rho(r)$ and electron fraction $Y_e(r)$ are taken from a one-dimensional (cold) neutron star model in hydrostatic equilibrium. The radius of 15 km is obtained for the Newtonian case. For a cool neutron star with baryonic mass of $M = 1.63\,M_\odot$ the general relativistic stellar structure equations yield a gravitational mass of about $M_g \approx 1.5\ M_\odot$ and a radius of 11.2 km with the equation of state of Lattimer & Swesty (1991). A softer equation of state than the one used in this work would imply a more compact star and a lower gravitational mass for the same baryonic mass.

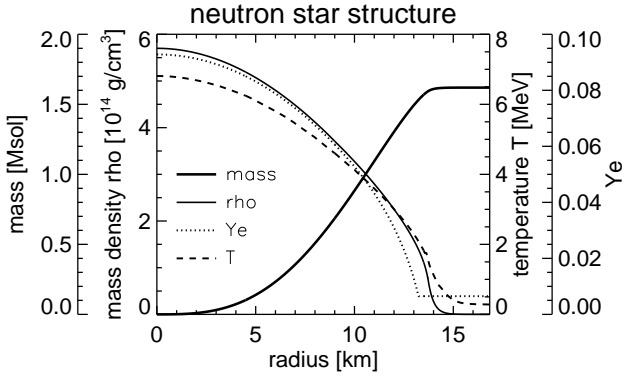

**Fig. 2.** Density, temperature, electron fraction, and enclosed baryonic mass as functions of radius for the initial neutron star model.

Note that we do not start our simulations with cold, $T = 0$, initial models. Due to the extreme sensitivity of the temperature to small variations of the internal energy in the degenerate limit (see Fig. 1), we decided to set the initial temperatures in our neutron stars to small, but finite values. By choosing temperatures corresponding to thermal energy densities of the order of 3% of the minimum internal energy density for a given density $\rho$ and electron fraction $Y_e$ (which is the degeneracy energy density at $T = 0$) we can avoid a great part of the temperature fluctuations. These are associated with the inversion of the equation of state for the temperature and are caused by small numerical errors in the internal energy as it is computed as the difference of total and kinetic energies in the PPM scheme. Fig. 2 shows density, temperature and electron number fraction of the initial neutron star model. The two-dimensional distribution of the initial temperature in the orbital plane can be seen in panels b of Figs. 4, 6, 14 and 16.

During the run, non-physical temperature spikes in isolated zones appear mainly due to the low spatial resolution combined with the highly degenerate state of the neutron star matter. These spikes also lead to non-conservation of energy. We reduce singular temperature spikes in each of the three axial directions by averaging over the two neighbouring zones. We do this only for zones in which the temperature was over 9 MeV, the density less than $1.5 \cdot 10^{14}$ g/cm$^3$, and the temperature contrast in both flanks of the spike are larger than 1 MeV. This procedure changes only the outermost surface layers of the neutron star and its negligible influence on the model evolution and the results was corroborated by the higher resolved model A128 in comparison with model A64.

The density of matter in the surroundings of the neutron stars is set to $10^9$ g/cm$^3$ with an internal energy equal to the value in the neutron star at the same density. To avoid this low-density matter being accelerated and falling onto the neutron stars, the velocities and kinetic energy are reduced to zero in zones in which the density is less than $3 \cdot 10^9$ g/cm$^3$. Additionally, the temperatures are also reset to their initial values where the density is smaller than $4 \cdot 10^{10}$ g/cm$^3$. These changes only affect relatively low-density material in the computational volume not filled by neutron star matter.

With a numerical grid of roughly 1 km spatial resolution one cannot accurately represent the density decline near the surface of the neutron stars, where the density typically decreases by a factor of ten (one dex) every 100 m over a distance of about 700 m. We artificially soften the edge by imposing a maximal change of 2 dex from zone to zone. The thickness of the surface layer thus results to 3 zones.

**Table 1.** Parameters and some computed quantities for all models. $N$ is the number of zones per dimension in the orbital plane, $S$ is the direction of the spin of the neutron stars, $T_{max}$ is the maximum temperature reached on the grid, $\widehat{\mathcal{L}}$ is the maximum gravitational-wave luminosity, $\mathcal{E}$ is the total energy emitted in gravitational waves, $\widehat{h}$ is the maximum amplitude of the gravitational waves at $r = 1$Gpc.

| Model | $N$ | $S$ | $T_{max}$ [MeV] | $\widehat{\mathcal{L}}$ [$10^{55} \frac{erg}{s}$] | $\mathcal{E}$ [$10^{52}$ erg] | $\widehat{rh}$ [$10^{-23}$] |
|---|---|---|---|---|---|---|
| A64 | 64 | 0 | 40 | 2.5 | 4.0 | 2.7 |
| B64 | 64 | +1 | 30 | 2.5 | 3.8 | 2.8 |
| C64 | 64 | −1 | >50 | 1.4 | 3.1 | 2.0 |
| A128 | 128 | 0 | 39 | 2.4 | 2.8 | – |
| T64 | 64 | 0 | – | 0.1 | – | – |

We prescribe the orbital velocities of the coalescing neutron stars according to the motions of point masses, as computed from the quadrupole formula. The tangential components of the velocities of the neutron star centers correspond to Kepler velocities on circular orbits, while the radial velocity components reflect the emission of gravitational waves leading to the inspiral of the orbiting bodies (see Kokkotas & Schäfer, 1995):

$$v_t = \omega r = \sqrt{\frac{GM}{2a_0}} \tag{34}$$

$$v_r = \dot{r} = -\frac{16}{5}\frac{G^3}{c^5}\frac{M^3}{a_0^3}\quad, \tag{35}$$

for $r = a_0/2$ and equal neutron star masses. Eq. 35 is valid in ADM coordinates (in harmonic coordinates the factor 16 is to be substituted by a factor 64). The initial angular frequency of the Newtonian Keplerian orbit is

$$\omega = \sqrt{\frac{2GM}{a_0^3}} \tag{36}$$

and the period follows to $P = 2\pi/\omega = 2.6$ ms for $a_0 = 42$ km and $M = 1.63 M_\odot$. Even if the neutron stars are born on very elliptic orbits, gravitational-wave emission drives the eccentricity towards zero before the neutron stars come close enough for mass exchange (e.g. Imshennik & Popov, 1994).

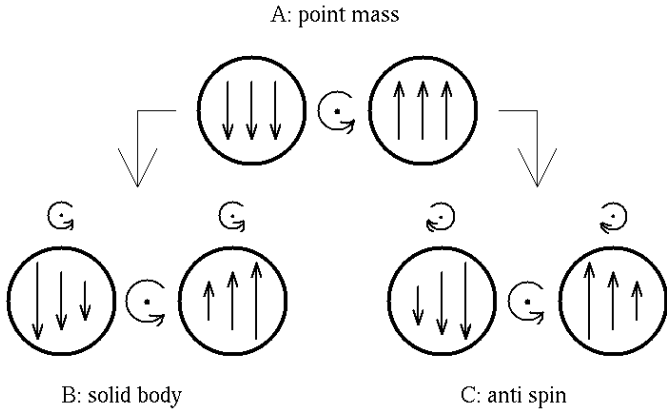

**Fig. 3.** A sketch of the three different initial spin distributions of models A, B, and C. The circular arrows between the neutron stars (circles) indicate the orbital motions, while the smaller circular arrows above show the additional spins of the neutron stars about their respective centers. The straight arrows inside the circles sketch the velocity distributions in the neutron stars: in model A all parts of the neutron stars are given the same velocity, while the additional spins in model B result in a solid body type rotation of the neutron stars. In model C the neutron stars are rotating opposite to the orbital spin direction.

A spin of the neutron stars around their respective centers is added and varied from model to model. Table 1 lists the parameters and some computed quantities for all models. The A models do not have any additional spin added on top of their orbital velocities. In this case all parts of the neutron stars start out with the same absolute value of the velocity. In models B64 and C64 a spin is added for each neutron star. The angular velocity (both magnitude and direction) of the spin in model B64 is equal to the angular velocity of the orbit. This results in a solid body type motion for model B64. In contrast, the direction of the angular velocity in model C64 is opposite to the orbit's angular velocity. The different cases are sketched in Fig. 3. We consider these three models to delimit in some sense extreme cases for the interaction of coalescing binary neutron stars. While the solid body case B64 represents a merging event with little internal shearing motions, model C64 is characterized by large velocities in opposite directions at the contact point of the neutron star surfaces and the coalescence is accompanied by violent shearing interaction. Model A128 has the same velocity distribution as model A64. However, it is computed with double the resolution in all dimensions and thus yields a test for the accuracy of the $64^3$ simulations. The two-dimensional velocity distributions in the orbital plane for models A64, A128, B64 and C64 can be seen in panels panels **a** of Figs. 4, 6, 14, and 16, respectively.

Model T64 is set up with the same initial conditions as model A64 except for the distance between the neutron stars which is chosen to be 60 km instead of 42 km (and the size of the grid consequently enlarged to 105 km a side). The larger initial separation of the neutron stars in model T64 corresponds to a time about 23 ms before two point masses reach the initial center-to-center distance of the neutron stars in models A64, A128, B64, and C64. This model is run for approximately 4.8 ms, which is about 1.1 orbital periods of two point masses at the given distance. Model T64 is intended to test whether the orbit is numerically stable for large initial separations. Moreover, the finite extension of the neutron stars leads to differences compared to the point-mass case and internal friction is checked in its effect on the temperature evolution during the spiral-in phase.

In Eq. 9 the term $+2P\frac{\partial v_i}{\partial x^i}$ can be replaced by the analytically equivalent term $-2v_i\frac{\partial P}{\partial x^i}$. This can be seen by applying the product rule to rewrite the derivatives and using the fact that surface terms are negligibly small. A test confirms that this replacement also does not produce any significant numerical difference in the resulting gravitational-wave luminosity $\mathcal{L}$ (Eq. 5). When model A64 is rerun for approximately 1.5 ms the difference of $\mathcal{L}$ between the two cases does not exceed 0.5% during this time.

The initial distance, at which the neutron stars are placed at the beginning of the simulation, is a compromise between physical accuracy and computational resources. On the one hand we would like to start with the neutron stars at a large distance in order to correctly simulate the tidal deformation they experience during inspiral. However, since the rate at which gravitational waves radiate energy away from the system increases with the fifth power of the distance (Eq. 24) the time to coalescence (which is proportional to the fourth power of the distance, Eq. 20) increases by a huge amount, too. The test calculations with an initial separation of the neutron stars of 60 km reveal a negligible orbital decay of about 1% radius decrease during 1 ms (one quarter orbit). Starting all simulations at such a large distance would be numerically prohibitive.

On the other hand, tidal deformation studies have been done (e.g. Lai et al, 1994; Reisenegger & Goldreich, 1994) which yield tidal deformations for extended objects of around 20% (for the principal axis) and normal mode excitation of at most a few percent of the radius at a distance of about 2.8 radii.

Two more test calculations clarify the influence of gravitational wave emission and of the dynamical tidal instability (Lai et al, 1994, and references cited therein). In one model the gravitational wave emission (and backreaction) is switched off and the initial orbit is taken to be circular. All the other initial conditions remain as described (in particular the separation of 42 km). In this case the neutron stars also merge, however on a longer timescale of approximately 2.5 revolutions. This merging is due to the fact that the two neutron stars are not in equilibrium in their common gravitational field at the beginning of the simulations and thus start to develop tidal deformations and oscillations which eventually bring them to the point of dynamical instability. Note that the initial separation of the two neutron stars (2.8 km) is only slightly larger than the minimum distance (2.6 km) for dynamical instability of objects *in* *equilibrium* (Lai et al, 1994). However, without gravitational wave emission the coalescence proceeds on a timescale that is 2–3 times longer. We therefore consider the approximation of spherical neutron stars instead of an initial equilibrium configuration as acceptable. In a second test model the gravitational wave emission (and backreaction) is again switched off, but the initial orbital velocities are given a radial component (Eq. 35) as if the preceding evolution had been influenced by gravitational wave emission. In this case the two neutron stars merge on nearly the same timescale as in the simulations that include gravitational radiation. The reason is the following. Since we start our calculations with the two neutron stars close to the distance where dynamical instability sets in, any additional radial velocity that decreases the orbital separation quickly pushes the neutron stars beyond the limit of dynamical instability.

Viscosity determines the velocity distribution within the neutron star: if it is large enough, spin-up during inspiral leads to tidal locking. However, Kochanek (1992), Bildsten & Cutler (1992) and Lai (1994) argue that it is very unlikely that microscopic shear and bulk viscosity alone are sufficient for spin-up or for tidal heating of the neutron stars to temperatures of more than $10^8$ K. For this reason, we consider model A that contains two non-corotating neutron stars as generic model. Models B and C represent variations obtained by adding spins to the individual neutron stars.

The calculations of models A64, B64 and C64 were performed on a Cray-YMP 4/64, needed about 16 MWords of main memory and took approximately 40 CPU-hours per model. Model A128 was computed on a Cray-EL98 4/256, required about 22 MWords of memory and 1700 CPU-hours.

## 4. Numerical results

### 4.1. Dynamic evolution

#### 4.1.1. Models A64, A128 and T64

We start our description of the hydrodynamic merging events with the models A64 and A128, which basically differ only in the resolution, model A128 having double the number of zones per dimension than model A64. Figs. 4 and 5 show the density, velocity, and temperature for the temporal evolution of model A64, and Figs. 6 and 7 give the corresponding information for model A128. The initial distance of the neutron stars in our models (approximately 2.8 neutron star radii) is very close to the separation where the configuration becomes dynamically unstable, which is approximately at a distance of 2.6 radii (Lai et al, 1994, and references cited therein). Therefore it is not surprising that already after one quarter of a revolution ($\approx 0.6$ ms, cf. Eq. 36) the neutron star surfaces touch due to the tidal deformation of the stars (Panels 4c and 6c). The parts of the neutron stars closest to the orbital axis experience the strongest tidal stretching and their temperatures decrease below 2 MeV (Figs. 4d and 6d), while the temperature of the bulk of the matter towards the geometric centers of the neutron stars falls from the initial 6 MeV to 4MeV.

As soon as the surfaces of the two neutron stars touch, the kinetic energy of the matter that moves in opposite directions is converted into internal energy and an elongated bar of hot material forms. Two fluid vortices appear at this time (Fig. 4e and 6e) and are associated with two separate hot "spots" (Fig. 4f and 6f). Isolated regions of high temperature can be observed during the whole subsequent evolution. Due to the strong gravity field of the massive, merging object the pressure varies smoothly within the merged object. Therefore regions with high temperature are associated with a lower density. This is visible in all plots when comparing the temperature and density contours, e.g. in Fig 6e and 6f.

Figure 8 shows the separation between the density maxima (*not* centers of mass) of the neutron stars. One notices, that for model A64 the neutron stars merge within 1.5 ms and then two density maxima form and separate again after about 2.5 ms. This swinging of the neutron stars, caused by their angular momentum, is not visible for model A128 in this figure. However, comparing the contours of the density in Figs. 5a and 7a, one recognizes that also in model A128 the two initial neutron stars can still be discriminated.

The evolution of the temperature distribution differs between models A64 and A128 after 4 ms as can be seen from a comparison of Figs. 5d and f with Figs. 7d and f, and is also clear from Fig. 9. This must partly be caused by the dissolving of the two fluid vortices, which are associated with the spots of high temperatures. In Fig. 5c the vortices are still present, while they are nearly completely dissolved and broken up into smaller ones in the better resolved model A128 at around the same time, Fig. 7c. The maximum temperature on the grid is shown in Fig. 9. The temperature increases by roughly 20 MeV for the A models when the surfaces of the two neutron stars come into contact.

However, because the matter is degenerate and the pressure not very sensitive to the temperature, the evolution of the density distribution is fairly similar in model A64 and model A128: Panels e in Figs. 5 and 7 show one merged object that contains most of the mass and is surrounded by a distended thick disk. In both models A64 and A128 the density contour corresponding to $\rho = 10^{11}$ g/cm$^3$ is approximately circular and extends nearly to the edges of the grid, i.e. has a diameter of approximately 80 km. The maximum density on the grid is displayed in Fig. 10 and is marginally higher for model A128 compared to the models with a resolution of 64 zones per dimension ($7.2 \cdot 10^{14}$ g/cm$^3$ compared to $6.9 \cdot 10^{14}$ g/cm$^3$).

The merged object has a total mass of approximately 3 M$_\odot$, which is larger than the maximum mass of stable neutron stars for the equation of state of Lattimer & Swesty (1991) in the general relativistic case. Therefore it is of interest to compare the size of the object with its Schwarzschild or event horizon. Since our computations are basically Newtonian it is not easy to estimate possible implications of a general relativistic treatment. A problem arises from the difficulty to choose the appropriate coordinate system for the discussion. We shall refer to harmonic coordinates or ADM (generalized isotropic) coordinates which are adequate for binary systems. Coordinate radii $\bar{r}$ in

**Fig. 4.** Contour plots of model **A64** showing cuts in the orbital plane for the density together with the velocity field (left panels) and for the temperature (right panels). The density contours are logarithmically spaced with intervals of 0.5 dex, while the temperature contours are linearly spaced with 2 MeV increments. The bold contours are labeled with their respective values, the density is measured in units of g/cm$^3$. The legend in each panel at the top right corner gives the scale of the velocity vectors and the time elapsed since the beginning of the simulation.

**Fig. 5.** Contour plots of model **A64** showing cuts in the orbital plane for the density together with the velocity field (left panels) and for the temperature (right panels). The density contours are logarithmically spaced with intervals of 0.5 dex, while the temperature contours are linearly spaced with 2 MeV increments. The bold contours are labeled with their respective values, the density is measured in units of g/cm$^3$. The legend in each panel at the top right corner gives the scale of the velocity vectors and the time elapsed since the beginning of the simulation.

**Fig. 6.** Contour plots of model **A128** showing cuts in the orbital plane for the density together with the velocity field (left panels) and for the temperature (right panels). The density contours are logarithmically spaced with intervals of 0.5 dex, while the temperature contours are linearly spaced with 2 MeV increments. The bold contours are labeled with their respective values, the density is measured in units of $g/cm^3$. The legend in each panel at the top right corner gives the scale of the velocity vectors and the time elapsed since the beginning of the simulation.

**Fig. 7.** Contour plots of model **A128** showing cuts in the orbital plane for the density together with the velocity field (left panels) and for the temperature (right panels). The density contours are logarithmically spaced with intervals of 0.5 dex, while the temperature contours are linearly spaced with 2 MeV increments. The bold contours are labeled with their respective values, the density is measured in units of $g/cm^3$. The legend in each panel at the top right corner gives the scale of the velocity vectors and the time elapsed since the beginning of the simulation.

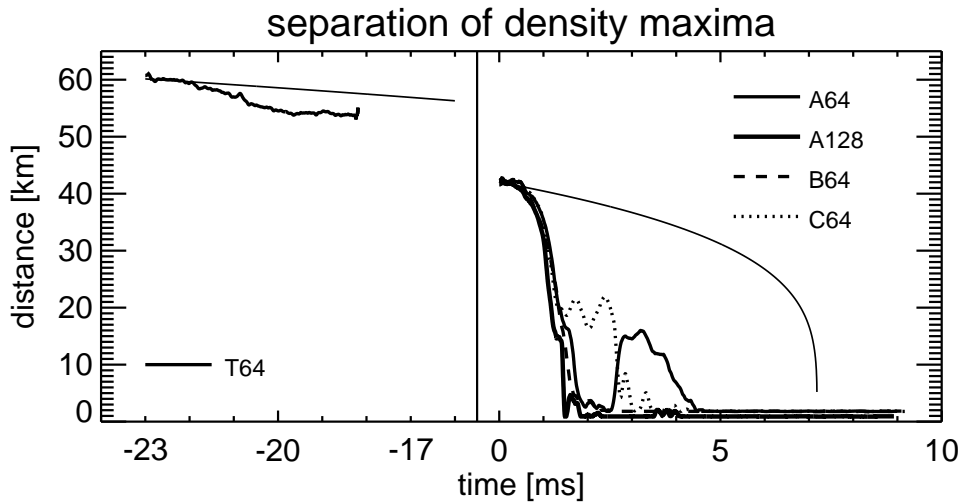

**Fig. 8.** The separation of the density maxima of the two neutron stars as a function of time for all five models. The thin, monotonously decreasing curve shows the distance between two inspiraling point-masses (Eq. 20). Note, that two point-masses at an initial distance equal to the center-to-center distance of the neutron stars of model T64 need about 23 ms to reach the initial separation of models A64, A128, B64, and C64.

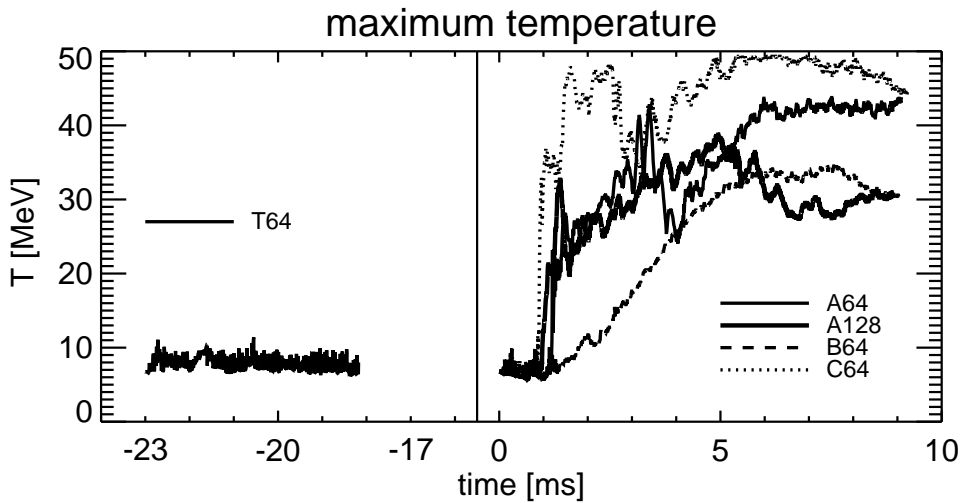

**Fig. 9.** The maximum temperature on the grid as a function of time for all five models.

Schwarzschild coordinates, $\tilde{r}$ in harmonic coordinates, and $r^*$ in isotropic coordinates are related by the following equation:

$$\bar{r} = \tilde{r} + 0.5\,R_s = r^*(1 + 0.25R_s/r^*)^2 \,, \tag{37}$$

where $R_s = 2GM/c^2$ is the Schwarzschildradius of a field generating mass $M$. The function $m(r)$ which gives the mass of the merged object enclosed by radius $r$ is shown in Fig. 11. For a given mass, the radius $r(m)$ at which the object attains this mass, is always *larger* than $0.5\ R_s(m(r))$, the radius of the Schwarzschild horizon in harmonic coordinates. ($R_s(m(r)) = 2Gm(r)/c^2$ is the Schwarzschildradius that corresponds to the enclosed mass $m(r)$.) However, for the mass range between about 1 and 3 $M_\odot$ the radius $3R_s$ lies *outside* the merged object.

$3R_s$ is, with an uncertainty of roughly 25%, suggested to be the radius of the last stable circular orbit for an equal-mass binary in harmonic coordinates (Kidder et al, 1992; Wex & Schäfer, 1993). This, of course, means that deeper investigations are needed. In Fig. 11 the straight line denoted by $2.5\ R_s$ gives the mass-radius relation in harmonic coordinates for the last stable circular orbit of a test particle with nonzero rest mass in a Schwarzschild field. In ADM coordinates the radius of the Schwarzschild horizon takes the value $0.25\ R_s$ and the radius of the last stable circular orbit of a nonzero-mass test body is about $2.475\ R_s$. The radius of $3.0\ R_s$ in harmonic coordinates is located at approximately $2.979\ R_s$ in ADM coordinates.

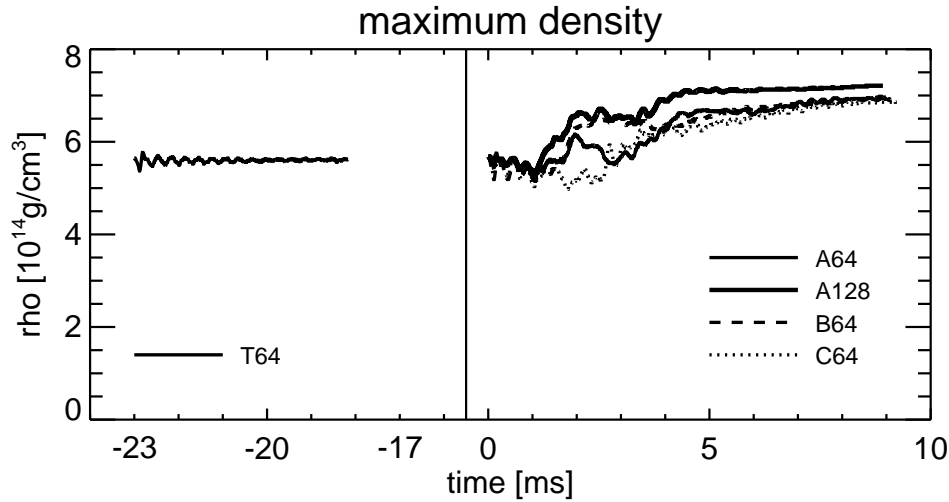

**Fig. 10.** The maximum density on the grid as a function of time for all five models.

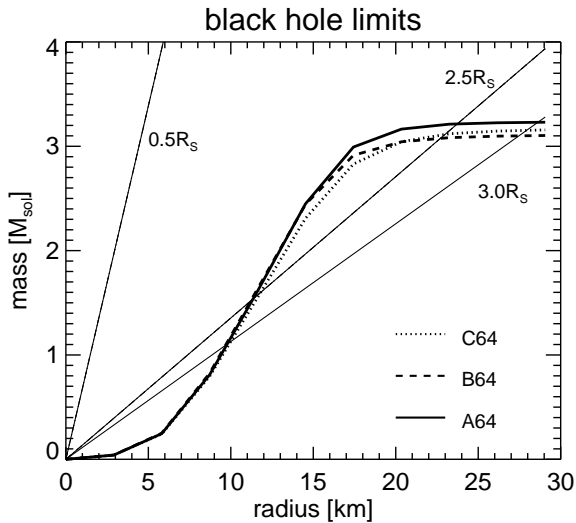

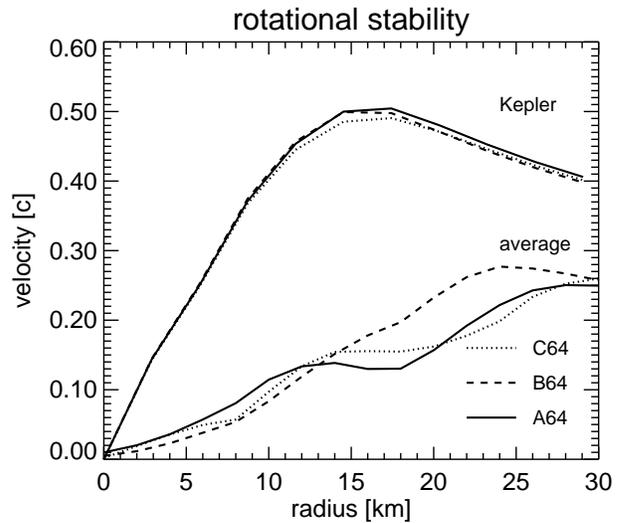

**Fig. 11.** The bold solid, dashed and dotted curves show the enclosed mass $m(r)$ as a function of radius (starting at the center of the coalesced object, i.e., at the grid center) at the end of the simulations. The three straight lines depict multiples of 0.5, 2.5, and 3.0 of the Schwarzschild radius $R_s(m(r))$ that corresponds to the mass on the vertical axis. In harmonic coordinates, 0.5 $R_s$ defines the radius of the Schwarzschild (event) horizon, 2.5 $R_s$ is the smallest radius of stable circular orbits for test particles with nonzero rest mass in a Schwarzschild field, and 3 $R_s$ denotes (roughly, not better than about 25%) the closest radius of stable circular orbits for an equal-mass binary.

**Fig. 12.** The upper curves show the Kepler velocity $v_k(m(r))$ that corresponds to the mass inside radius $r$ as seen in Fig. 11. The lower curves depict the average orbital velocity in the merged objects as function of the distance from the grid center.

factor of 4 smaller than the Kepler velocities, so centrifugal forces are not expected to be strong enough to prevent or delay the collapse to a black hole.

The density and temperature distributions of the test model T64 can be found in Fig. 13 for two snapshots in time. The first two panels a and b show the initial state, while the panels c and d show the distributions after about 1.2 orbital revolutions. Only relatively minor changes of the structure occur during this time: the thickness of the surface layers varies, the shapes of the neutron stars are slighly deformed from the original spheres, and the internal temperatures in the stars de-

When the merged object rotates so fast that centrifugal forces are important the collapse to a black hole might be delayed by the stabilizing effect of these forces. Fig. 12 compares the Kepler velocity at a given radius, which is only dependent on the mass inside that radius, to the average model velocities at the same radius. One can see that the actual velocities are some

**Fig. 13.** Contour plots of model **T64** showing cuts in the orbital plane for the density together with the velocity field (left panels) and for the temperature (right panels). The density contours are logarithmically spaced with intervals of 0.5 dex, while the temperature contours are linearly spaced with 2 MeV increments. The bold contours are labeled with their respective values, the density is measured in units of $g/cm^3$. The legend in each panel at the top right corner gives the scale of the velocity vectors and the time elapsed since the beginning of the simulation.

crease due to the net expansion and increase of the volumes. The maximum density (shown in Fig. 10) oscillates with a period of approximately 0.31 ms (which is the sound crossing time of the neutron stars) around an average value of $5.6 \cdot 10^{14}$ $g/cm^3$. This oscillation has a damping timescale of approximately 2 ms. Note that the maximum density of the initial neutron star model is $5.7 \cdot 10^{14}$ $g/cm^3$. The maximum temperature fluctuates around 8 MeV (Fig. 9), but these temperatures affect only few zones none of which appear in Fig. 13. These results for test model T64 show us that it is rather unlikely that the evolution preceding the initial configurations of our models A64,

A128, B64, and C64 will produce strongly distorted and heated neutron stars.

### 4.1.2. Models B64 and C64

Model B64 (Figs. 14 and 15) represents the evolution of two neutron stars that rotate like one rigidly rotating solid body. The corresponding initial velocity distribution was adopted in a number of other investigations, e.g. in most of the models computed by Shibata et al (1993) and Nakamura & Oohara (1991), and in earlier works of this group. The assumption of rigid rotation appears natural when tidal locking is assumed due to the action of viscous forces. This situation, however, does not appear to be likely (see discussion in Sect. 3).

**Fig. 14.** Contour plots of model **B64** showing cuts in the orbital plane for the density together with the velocity field (left panels) and for the temperature (right panels). The density contours are logarithmically spaced with intervals of 0.5 dex, while the temperature contours are linearly spaced with 2 MeV increments. The bold contours are labeled with their respective values, the density is measured in units of g/cm$^3$. The legend in each panel at the top right corner gives the scale of the velocity vectors and the time elapsed since the beginning of the simulation.

**Fig. 15.** Contour plots of model **B64** showing cuts in the orbital plane for the density together with the velocity field (left panels) and for the temperature (right panels). The density contours are logarithmically spaced with intervals of 0.5 dex, while the temperature contours are linearly spaced with 2 MeV increments. The bold contours are labeled with their respective values, the density is measured in units of $g/cm^3$. The legend in each panel at the top right corner gives the scale of the velocity vectors and the time elapsed since the beginning of the simulation.

**Fig. 16.** Contour plots of model **C64** showing cuts in the orbital plane for the density together with the velocity field (left panels) and for the temperature (right panels). The density contours are logarithmically spaced with intervals of 0.5 dex, while the temperature contours are linearly spaced with 2 MeV increments. The bold contours are labeled with their respective values, the density is measured in units of $g/cm^3$. The legend in each panel at the top right corner gives the scale of the velocity vectors and the time elapsed since the beginning of the simulation.

**Fig. 17.** Contour plots of model **C64** showing cuts in the orbital plane for the density together with the velocity field (left panels) and for the temperature (right panels). The density contours are logarithmically spaced with intervals of 0.5 dex, while the temperature contours are linearly spaced with 2 MeV increments. The bold contours are labeled with their respective values, the density is measured in units of $g/cm^3$. The legend in each panel at the top right corner gives the scale of the velocity vectors and the time elapsed since the beginning of the simulation.

Model B64 shows important differences in the initial velocity distribution when compared with model A64. The velocities at the neutron star surface regions that face each other are much smaller in model B64. When these surfaces touch, the dissipation of kinetic energy into heat is much weaker. This can be seen very clearly by comparing Fig. 14f with Fig. 4f. While in model A64 the temperatures have reached 14 MeV after about 1.7 ms, they are only 8 MeV in model B64. This difference is also visible in the maximum temperatures (Fig. 9), where model B64 is the only model (apart from the test model T64) that does not show a sudden temperature jump at approximately 1 ms. Although eventually also in model B64 two hot spots develop (Fig. 15d), these are less prominent and dissolve towards the end of the simulation (Fig. 15f) and form a ring of hot material in the orbital plane. Again (cf. model A128), this dissolution happens at the same time when the vortex motions disappear.

Differences of the velocity distribution of models A64 and B64 also exist at those parts of the surfaces of the neutron stars that are most distant from the orbital rotation axis. Since the velocities of model B64 are much larger here, the centrifugal forces are also much stronger, which in turn facilitates mass shedding from the outer layers. Thus, two prominent spiral arms form (Fig. 15a) from which matter is lost off the grid. Fig. 18 shows the amount of gas that flows off the grid as a function of time for all models. Model B64 sheds approximately five times more matter than model A64, and model A128 is another factor of two below model A64. Although initially prominent, the spiral arms (on our grid) do not seem to survive several revolutions and after about 4.5 ms a thick disk, just as in model A64, has developed. Fig. 15e does not differ qualitatively from Fig 5e or Fig 7e.

Most of the mass that flows off the grid remains bound to the system. Fig. 19 shows the cumulative amount of matter that gets unbound as a function of time. For every zone along the grid boundaries, we check whether the sum of potential, kinetic, and internal energy of the outward streaming gas is positive. To estimate the amount of unbound matter we add up the mass flow across the boundaries only for these zones. We find that a few times $10^{-4} M_\odot$ might be able to escape from the system. These numbers are obtained by using the thermal energy of the gas which is the differerence between its internal energy and its internal energy at $T = 0$ for given values of density and $Y_e$. Taking the total internal energy instead of just the thermal energy could lead to somewhat larger estimates of the mass that gets gravitationally unbound. In case of model B64 the difference is about a factor of 3 (Fig. 19). Neutrino-energy deposition in the cool outer regions of the disk could power a low-density mass flow like the neutrino-driven wind from a newly formed neutron star in a Type II supernova. Moreover, energy release from the recombination of nucleons or from nuclear reactions (Davies et al, 1994) could have dynamical effects and could increase the mass loss from the disk. All these processes suggest that our numbers for the unbound mass are only lower estimates. The stronger gravitational potential in a fully general relativistic treatment, on the other hand, would imply a stronger binding of the disk material and would lead to a corresponding reduction of the mass loss.

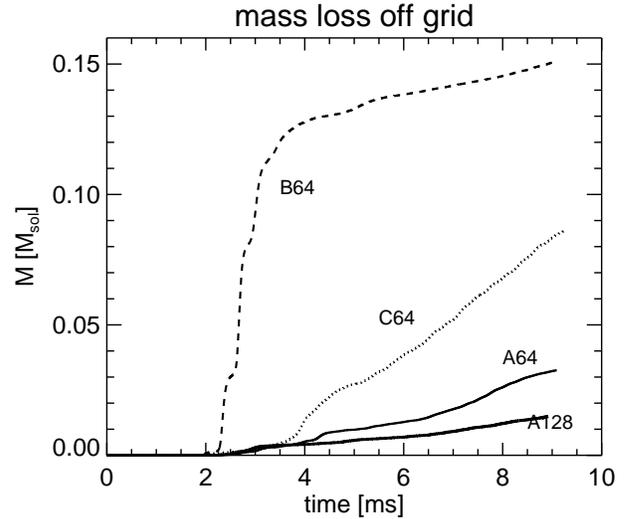

**Fig. 18.** The cumulative amount of matter that flows off the grid during the simulations as a function of time for models A64, A128, B64, and C64.

If one interprets model B64 as emerging from model A64 by adding spins to the neutron stars around their respective axes, then model C64 is constructed from model A64 by adding negative spins. Thus, contrary to model B64, the velocities in C64 are largest at points closest to the orbital rotation axis (Fig. 16a). The differences to model A64 as discussed for model B64 above are now found to be present in the reverse direction. As soon as the neutron star surfaces touch each other the temperatures quickly rise to high values, roughly 30 MeV (Fig. 16f), and later to even more than 45 MeV (Fig. 9). As already seen in model A64, the hot contact region of the neutron stars splits into two isolated hot spots which remain present until the end of the simulation (Fig. 17b, d, and f).

The high velocities at the neutron star surfaces facing each other also lead to a stronger deformation of the approaching stars in model C64 (Fig. 16c) compared to models A64 (Fig. 4c) and B64 (Fig. 14c), the latter two developing ellipsoidal deformations with the major axes of the neutron stars being aligned shortly before coalescence. On the other hand, the velocities of the parts of the neutron star surfaces facing away from the orbital rotation axis are small and no pronounced spiral arm extensions like in model B64 appear (compare Fig. 16e with Fig. 14e and Fig. 17a with Fig. 15a).

At the end of our simulations (Figs. 5e,f, 15e,f, 17e,f) the coalesced object has the following structure. Most of the mass (96%) is contained in a high-density central, rotating body that is slightly triaxial. The major axes of the isodensity contours are not aligned. The central body wobbles and rings and continuously sends pressure and density waves into the surrounding, thick, distended disk which is formed by the matter in the outer parts of our computational grid. These waves transfer angular momentum and influence the disk's structure and extension, in particular cause periodic expansion and contraction motions with spiral-wave-like appearance. Moreover, the interactions of the waves generates heating of the disk gas over the simulated period of several milliseconds after the merging of the neutron stars. Probably also most of the material that flows off the grid (about 0.1 $M_\odot$) would add to the disk, because we expect only very little (a few $10^{-4} M_\odot$) of this matter to get unbound. Maximal temperatures occur in hot spots located in a ring of material at a distance of about 10–15 km around the central density maximum. The densities in this very hot layer are typically 1–2$\cdot 10^{14}$ g/cm$^3$.

**Fig. 20.** Contour plots of the four models A64, A128, B64, and C64 showing the density distribution in cut planes perpendicular to the orbital plane. The density contours are logarithmically spaced with intervals of 0.5 dex. The bold contours are labeled with their respective values. The times at which these snapshots are taken are the same as the times of panels **f** of Figs. 5, 7, 15, and 17, respectively.

**Fig. 21.** Contour plots of the four models A64, A128, B64, and C64 showing the temperature distribution in cut planes perpendicular to the orbital plane. The temperature contours are linearly spaced with 2 MeV increments. The bold contours are labeled with their respective values. The times at which these snapshots are taken are the same as the times of panels **f** of Figs. 5, 7, 15, and 17, respectively.

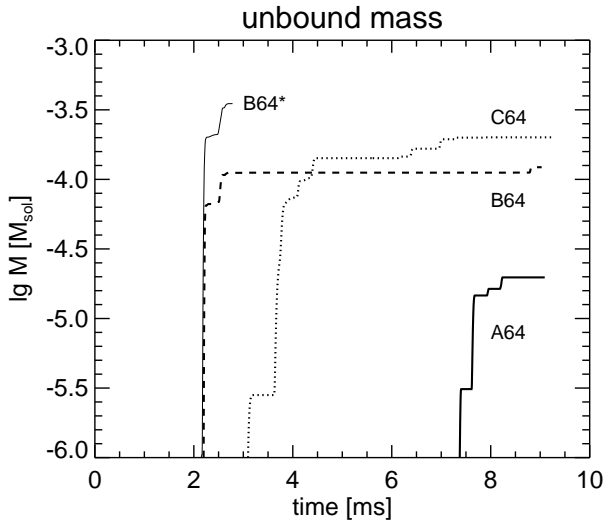

**Fig. 19.** The cumulative amount of matter that is unbound when flowing off the grid as a function of time for models A64, B64 and C64. The thick dashed, dotted, and solid lines, respectively, are computed with the criterion that the sum of gravitational, kinetic, and thermal energy of the gas are positive. In this case no matter gets unbound in the better resolved model A128. Using the total internal energy instead of only the thermal energy leads to the result given by the thin solid line for model B64 (denoted by B64*).

Density and temperature distributions for models A64, A128, B64, and C64 in two orthogonal planes perpendicular to the orbital plane are shown in Figs. 20 and 21. The displayed snapshots correspond to the times of the last panels **f** of Figs. 5, 7, 15, and 17, respectively. The high-density cores are essentially spherically shaped in all of the cases of Fig. 20, only in model A128 small cuspy bulges of the $\rho = 10^{14}$ g/cm$^3$ contour at the equator (orbital plane) indicate the very fast rotation of the massive central object. Also, model A128 has steeper density gradients at the poles than model A64, which suggests that the vertical extension of the disk and its shape are somewhat dependent on the numerical resolution. With higher numerical resolu-tion one should expect a more compact and more elliptically deformed disk structure vertical to the orbital plane. Asymmetries of the disk to both sides of the center and differences between the two cut directions result from the perturbations and waves sent into the surrounding gas by the dynamical contractions and expansions of the central object. The temperature distributions of Fig. 21 reveal a rather complex structure. The ring-like region of highest temperatures ($T \gtrsim 15$–20 MeV) around the cooler central core (see above) has a crescent or banana-shaped cross section. Hot knots outside the equatorial plane can occur and are ordered in a quadrupole type manner. In all models the regions at the poles remain comparatively cool with temperatures similar to those present in the inner core ($T \lesssim 15$–20 MeV). The outer parts of the disk have temperatures around 2–4 MeV in regions where the density is a few $10^{11}$ g/cm$^3$ or less.

Since our calculations are Newtonian, the merged object with a mass of about 3 $M_\odot$ remains stable and does not collapse into a black hole. However, in a general relativistic treatment gravitational instability would be unavoidable on a timescale of a few milliseconds from the start of our modelling, because the mass of 3 $M_\odot$ is beyond the maximum stable neutron star mass (about 2.2 $M_\odot$) associated with the used equation of state.

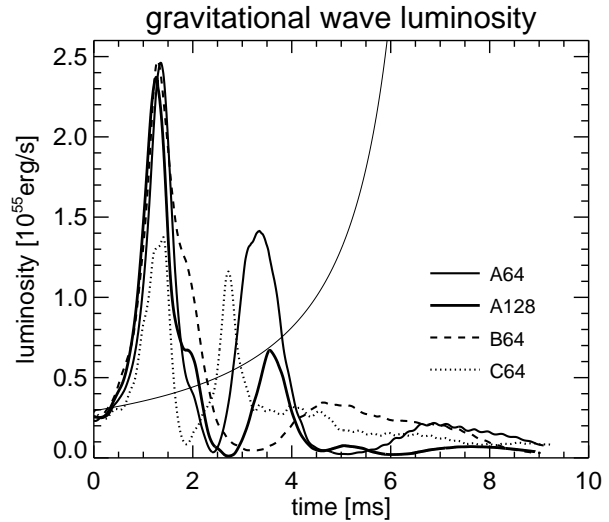

**Fig. 22.** The total emission rate of gravitational wave energy as a function of time for models A64, A128, B64, and C64. The monotonously rising curve shows the emission for a point-mass binary (Eq. 24).

### 4.2. Gravitational waves

The gravitational wave luminosity (Eq. 5) is shown as a function of time in Fig. 22. Initially, the luminosity of our models is nearly equal to the emission of binary point-masses. After about 1 ms the dynamical instability sets in and the neutron stars merge. Therefore, within less than 1.5 ms the power of the models reaches a very high, narrow maximum. Model A has a second, less pronounced luminosity peak at about 3.5 ms, while model B shows a second, broad hump around 5 ms. Model C is exceptional because its first luminosity maximum is only roughly half as high as that of the other models and is therefore only slightly stronger than the following, second maximum at about 2.75 ms.

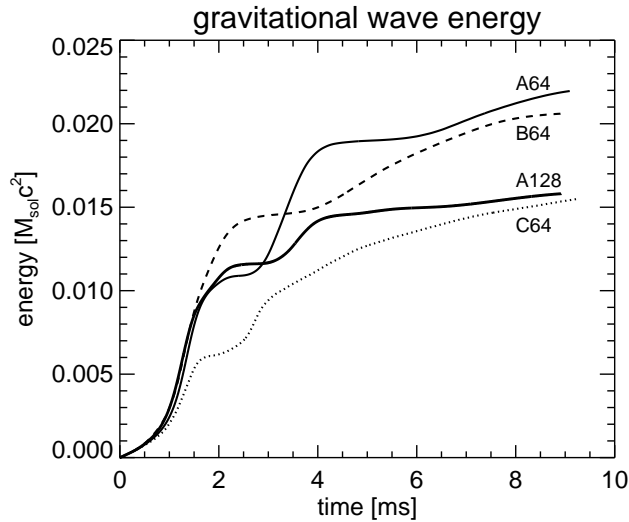

**Fig. 23.** The integrated energy emitted in gravitational waves as a function of time for models A64, A128, B64, and C64.

There seems to be a trend from model B over A to C, associated with the decrease of the total angular momentum (see Fig. 3): the relative height of the first to the second luminosity maximum decreases.

The time integral of the luminosities, i.e. the energy emitted in gravitational waves, is shown in Fig. 23. Although the main phases of emission occur between 1 and 4 ms, a slow rise of the curves towards the end of the simulations indicates that even during the late stages of the simulated evolution a considerable amount of energy continues to be emitted. The total energy radiated by our models during the computed neutron star coalescence and subsequent evolution lies between 1.5% and 2.2% of $M_\odot c^2$.

In Figs. 22 and 23 differences of the gravitational wave emission between models A64 and A128 are clearly visible. Model A64 has a much stronger second luminosity peak and, correspondingly, its total energy emitted in gravitational waves is roughly 40% higher. These discrepancies are most likely a consequence of the coarser numerical resolution of model A64. Although the structure of the PPM code explicitly conserves energy and momentum (in the absence of a gravitational potential term), this does not hold true for the angular momentum. During the early, very dynamical phase of the merging of the neutron stars, fluid whirls and vortices form which contain a significant fraction of the kinetic energy and angular momentum of the initial orbital motion. The distribution and amount of angular momentum plays an essential role for the subsequent behavior and structure of the merged object. This is obvious from the comparison of models A64, B64, and C64. During the merging process minor deviations between models A64 and A128 are already visible at $t \approx 0.7$ ms (Figs. 4c,d and 6c,d), increasing differences at $t \approx 1.7$ ms (Figs. 4e,f and 6e,f), and significant ones at $t \approx 3.1$–3.3 ms (Figs. 5a,b and 7a,b). In model A128 structures and density and temperature gradients are not only sharper and better represented, but, most important, the fluid motion appears less coherent and the flow pattern decays into a larger number of smaller vortices during the following evolution.

These differences of the fluid flow are reflected in the time evolution of the separation of local density maxima as displayed in Fig. 8. The coherence of the flow in model A64 is associated with a sizable amount of angular momentum in the large-scale flow and leads to a transient re-formation of two distinct density maxima after about 2.5 ms and after an intermediate period where the neutron stars had already merged into a single, very compact body. In contrast, in model A128 much angular momentum is located in the smaller and more fine-stuctured vortices and the angular momentum in the overall rotation is not sufficient to support a partial re-separation of the merged neutron stars after coalescence. This is apparent in Figs. 5a and 7a where the high-density, central body of model A64 shows a very elongated shape while in model A128 it is more compact and spherical. The discussed dynamical consequences of the numerical resolution of both models are also responsible for the differences between A64 and A128 seen in Figs. 9 and 10. Moreover, there is a positive correlation between the gravitational wave luminosity (Fig. 22) and the separation of the density maxima in the merging system (Fig. 8): the gravitational wave emission of models A64 and A128 is very similar up to about 2.5 ms which is the time when the two density maxima of the neutron stars in model A64 start to reappear. Between 2.5 ms and about 4.5 ms A64 shows its pronounced second luminosity peak, which perfectly corresponds to the duration of the intermediate phase of the re-formation of two distinct, local density maxima. After 4.5 ms both models A64 and A128 have developed a nearly spherical high-density core and their gravitational wave emission becomes rather weak.

A detailed evaluation of our models reveals that during the first 1–2 ms, which is the phase of the violent, dynamical merging (see Fig. 8), model A64 loses about 10% of its initial angular momentum, while in the case of A128 the violation of angular momentum conservation seems to be negligible before about 4.5 ms. Only during the subsequent evolution ($t \gtrsim 4.5$ ms) the cumulative loss of angular momentum grows roughly linearly with time in A128 to reach a few per cent at the end of our simulation. This decrease of the angular momentum in the computational volume is primarily caused by the flow of matter off the grid, see Fig. 18, which occurs at a constant rate during the late evolution of model A128. Model A128 seems to be well resolved and angular momentum conserved to a high degree of accuracy. Since gravitational waves do not only carry away energy but also angular momentum it is necessary to distinguish carefully between the physical losses mentioned above and the numerical destruction of angular momentum. To check our analysis we compare with a simulation of the dynamical coalescence where the effects of gravitational waves are completely switched off. In case of a computation with $64^3$ grid points the angular momentum is satisfactorily conserved until about 0.7 ms but between 1 and 2 ms about 6% of the angular momentum are destroyed. Although the merging proceeds in a different way when gravitational waves are disregarded, the similarity of the numbers suggests that our conclusions about the numerical loss of angular momentum should be correct.

The gravitational waveforms that correspond to these luminosities are shown in Figs. 24 and 25. In both figures the results of the numerical models are plotted by bold lines, while the analytic results for the point-mass binary are plotted by thin lines. For model T64 the time $t \approx -23$ ms coincides with the beginning of the simulations. The finite extension and deformation of the neutron stars in the test model T64 produces noticable deviations from the point-mass solution after approximately one revolution (Fig. 25). A part of this effect, however, may also be caused by the low resolution of this model with only 9 mesh zones on the length of one neutron star radius.

The amplitude and frequency of the waves in the point-mass approximation increases monotonously until the two point masses finally meet at $t \approx 7$ ms (Fig. 24). For an initial phase of about 1 ms the waveforms of the numerical models A, B, and C coincide well with those of the point-mass binary. As soon as the neutron stars get tidally deformed, the waveforms deviate in amplitude and frequency. The merged neutron stars tend to produce waves at a dominant frequency between 1.5 and 2 KHz as can be seen in Fig. 26 which gives the Fourier transform of the gravitational waveforms of Fig. 24. Beyond the broad maximum the power spectra of the gravitational wave emission show a steep decline towards higher frequencies with an average power-law index of about $-7....-10$.

The cumulative emission of gravitational-wave energy as a function of frequency is shown in Fig. 27. In the upper part of each panel, the downward sloping straight line represents the energy loss per unit frequency interval of a point-mass binary as given by Eq. 26. The frequencies to the left of the vertical line correspond to the wave frequencies that are emitted before the start of the numerical modelling, i.e. for times $t < 0$ (see Fig. 24). We produce a combined wave by using the quadrupole moments for a point-mass binary for $t < 0$ and taking the numerically obtained quadrupole moments (Eq. 19) at $t > 0$ (see Zhuge et al, 1994). The combined wave is then Fourier analysed up to the times given in the lower left corners of the panels of Fig. 27. With the Fourier transform, the energy spectrum is calculated via Eq. 25.

At low frequencies the energy spectrum calculated for the combined waves fits very well to what is expected from the point-mass approximation. Although the merged object in the simulations radiates gravitational waves for a longer time than the point-masses, the very much smaller amplitudes (Fig. 24) result in less energy emitted at almost all higher frequencies. A prominent peak is visible for all models at about 2 KHz. By Fourier transforming the signal of the combined wave up to different times $t > 0$, we are able to roughly locate the time at which the peaks are produced. At around 1.6 ms all models have a featureless, monotonously declining spectrum at high frequencies. Model A64 starts to produce its 2 KHz maximum at around 3 ms, while model B64 shows indications of its maximum after about 5 ms. Model C64 has developed a prominent 2 KHz peak already at 3 ms.

**Fig. 24.** The gravitational waveforms, $h_+(0,0)$ and $h_\times(0,0)$, for the models A64, B64, and C64 are plotted by bold, solid lines beginning at $t = 0$. The thin, solid lines represent the analytic result for a point-mass binary in the quadrupole approximation (Eqs. 21 and 22).

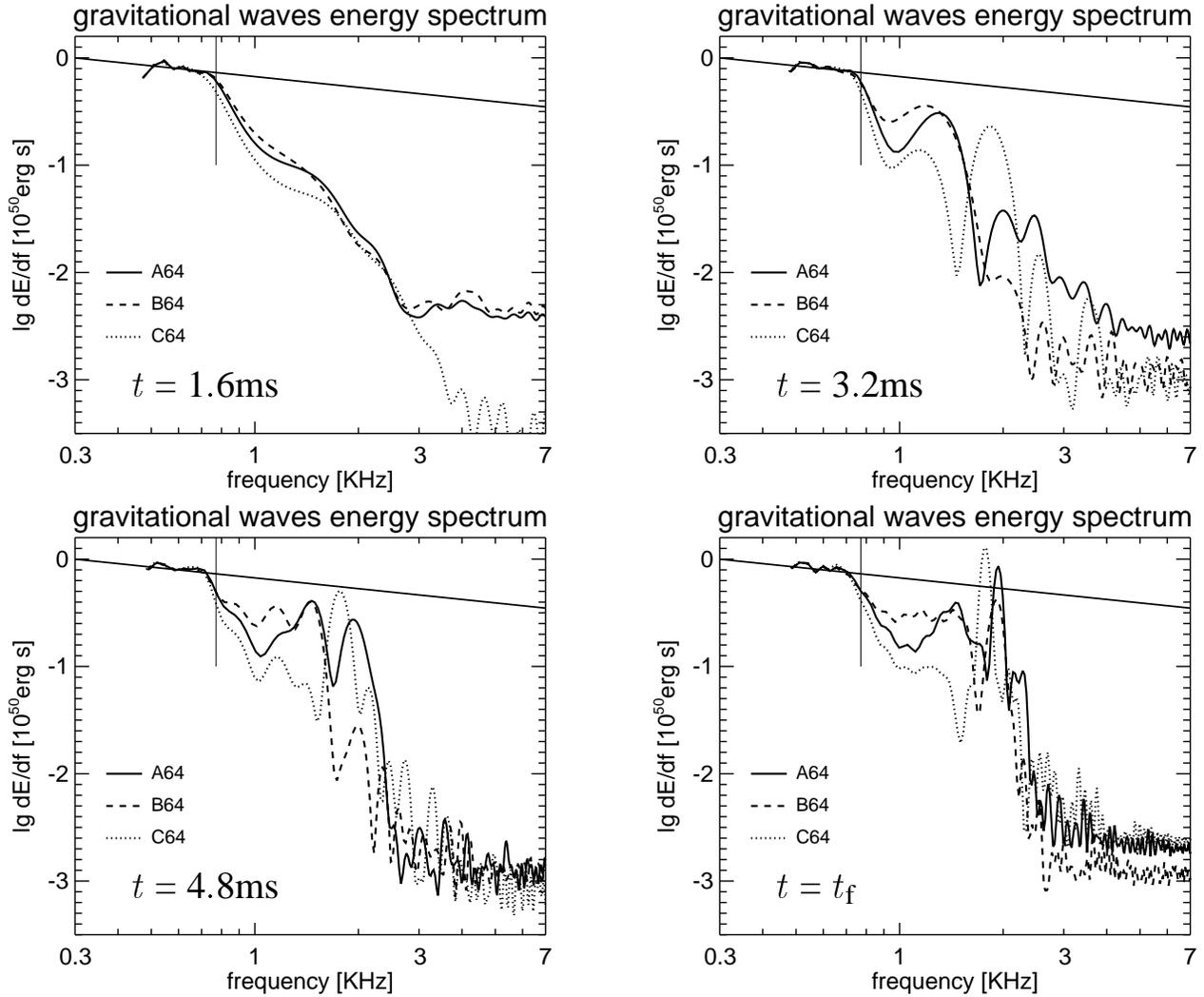

**Fig. 27.** Snapshots of the energy spectrum of emitted gravitational waves at four different times for the models A64, B64, and C64. The time up to which the Fourier transform is performed is indicated in the lower left corner of each panel. The straight, downward sloping line is the spectrum of a point-mass binary, as given by Eq. 26. The vertical lines indicate the frequency corresponding to the orbital frequency of two point masses at the initial distance of the neutron stars in our numerical models.

The gravitational waves yield information about the acceleration of the tensor components of the mass quadrupole moment of our system or, according to Eq. 19, about the trace-free part of the sum of two times the system's kinetic energy tensor and the stress tensor for the Newtonian gravitational field. As is well known, taking the trace-free part is equivalent to subtracting the spherically symmetric part. Thus, the gravitational wave field gives only information about non-spherically symmetric (stress-energy) aspects of its source.

While gravitational waves picture the (hydro-)dynamics and mass motions and are produced in the KHz regime, the radiation of neutrinos, which are emitted with thermal energies typical of the region around the neutrino sphere, reflects the thermo-dynamical properties of the emitting matter, in particular of the disk material that surrounds the massive central object formed after the merging of the binary neutron stars. The characteristics of the neutrino emission and the corresponding implications for gamma-ray burst models will be addressed in a forthcoming paper.

## 5. Summary and discussion

The hydrodynamic simulations presented in this work follow the density and temperature evolution as well as the gravitational wave emission of two merging neutron stars. Initially, each of the two identical, cool neutron stars is in hydrostatic equilibrium and has a baryonic mass of $\approx 1.6\,M_\odot$ and a radius of $\approx 15$ km. The neutron stars are placed at an initial center-to-center distance of 42 km. No attempt is made to construct

## waveform, model T64

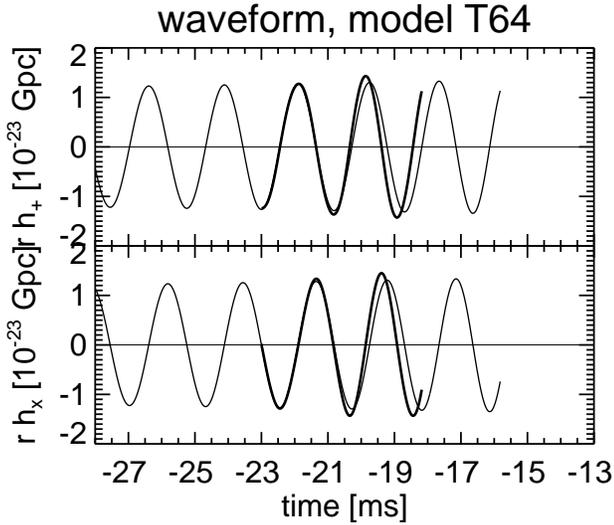

**Fig. 25.** The gravitational waveforms, $h_+(0,0)$ and $h_\times(0,0)$, for the test model T64 are plotted by bold, solid lines beginning at $t = -23$ ms. The thin, solid lines represent the analytic result for a point-mass binary in the quadrupole approximation (Eqs. 21 and 22).

## gravitational waves power spectrum

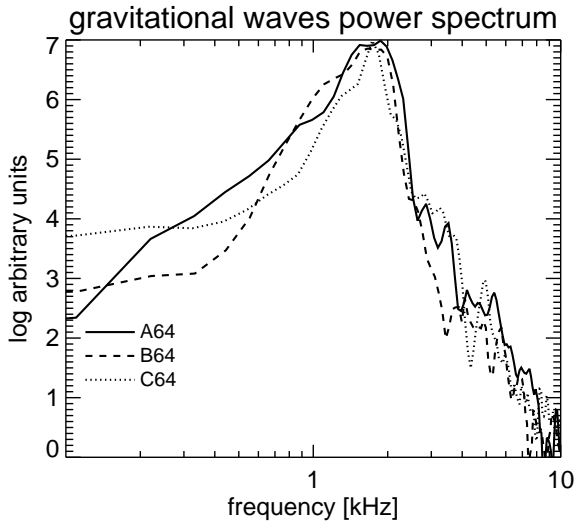

**Fig. 26.** Power spectrum (Fourier transform) of the gravitational waveforms shown in Fig. 24 for the models A64, B64, and C64.

initial equilibrium configurations of the binary system. The initial temperatures in the neutron stars are increased above the cold, $T = 0$, situation in such a way that the thermal energy is about 3% of the degeneracy energy for given density and electron fraction. This yields a central temperature of about 8 MeV at the beginning of our simulations. The orbital velocities of the coalescing neutron stars are prescribed according to the motions of point masses as computed from the quadrupole formula. Spins are added to the neutron stars to take into account rotations of the neutron stars around their axes vertical to the orbital plane. The spins are different from model to model. We simulate the three cases without neutron star spins and with spin vectors parallel and antiparallel to the vector of the orbital angular momentum. The case with neutron star spins parallel to the orbital spin yields a rigid solid-body rotation. Recent work (Kochanek, 1992; Lai, 1994; Reisenegger & Goldreich, 1994) suggests such a small viscosity of neutron star matter that the stars cannot develop synchronous rotation during inspiral. This gives us the freedom to set up initial spins without constraints from the orbital motion. Contrary to Rasio & Shapiro (1994) who considered the corotating case only, we therefore do not choose our initial conditions as exact equilibrium configurations.

The orbit decays due to gravitational wave emission, and after less than one revolution the stars are so close that dynamical instability (Lai et al, 1994, and references cited therein) sets in. Within 1 ms they merge into a rapidly spinning ($P_{\rm spin} \approx 1$ ms), high-density body ($\rho \gtrsim 10^{14}$ g/cm³) with a surrounding thick disk of material with densities $\rho \approx 10^{10} - 10^{12}$ g/cm³ and rotational velocities of 0.3–0.5 c. In our Newtonian models a merged object with a mass of approximately 3 M$_\odot$ is formed. This mass is far above the limiting mass of relativistic neutron stars for the used equation of state of Lattimer & Swesty (1991) and is also in excess of the limiting masses of most currently discussed supranuclear equations of state. We therefore expect that the central object in our simulations would collapse into a black hole on a time scale not much longer than the dynamical timescale of approximately 1 ms, if the computations were performed with a fully relativistic treatment. The duration of the stable phase prior to collapse will depend on the masses of the neutron stars and on the properties of the equation of state. The stability of the coalesced object in the context of uncertainties of the supranuclear equation of state was discussed in more detail by Davies et al (1994) and the possible effects of additional support of the merged object by centrifugal forces associated with the large angular momentum was addressed by Rasio & Shapiro (1992) and Davies et al (1994).

Because of the different angular momentum in the different models, the amount of matter that flows off the computational grid varies between 0.02 and 0.15 $M_\odot$. Since this matter has a high specific angular momentum it will most likely add to the disk around the merged object and might even remain in the disk when the massive central object has collapsed into a black hole. At the moment when crossing the grid boundaries, only a fairly small fraction of this matter has the energy to escape from the gravitational potential of the merged binary. Our estimates yield a few $10^{-4}$ $M_\odot$ of unbound material. Yet, momentum and energy transfer by pressure waves, neutrino heating, and nuclear energy release might lead to additional mass being stripped off the outer regions of the disk and could increase the mass loss. The use of a grid-based code like PPM does not allow us to follow the evolution of the tidally-formed spiral arms which reach far out from the merged neutron stars and are very prominent in the SPH simulations of Rasio & Shapiro (1994), Davies et al (1994), and Zhuge et al (1994). The formation of these spiral arms, however, is visible in our model with solid-body like ro-tation. This model accordingly also shows the maximal amount of mass-shedding off the grid ($0.15\ M_\odot$). The presence of spiral arms clearly depends on the neutron star spins and the corresponding angular momentum in addition to the orbital angular momentum. This conclusion can also be drawn on grounds of the results and figures in Shibata et al (1992). The dilute disk extends to the grid boundaries not only in the orbital plane but reaches out to the boundaries also in the vertical direction. A steep density gradient towards the edges of our grid and comparatively high densities ($10^{10}$–$10^{11}\ \mathrm{g/cm^3}$) indicate that the thickness of the disk certainly exceeds 40 km which is the vertical height of our computational volume. Davies et al (1994) find a butterfly-shaped cross section of the disk perpendicular to the plane of the neutron star trajectories in their SPH simulations.

At the end of our simulations when the merger has reached a state near rotational equilibrium, the central, compact object formed after the coalescence of the neutron stars is essentially spherical for the model with the lowest total angular momentum (model C64). In case of the other models slight triaxiality occurs with major axis ratios of the density contour for $\rho = 10^{14}\ \mathrm{g/cm^3}$ of about $1.15 : 1$ in the orbital plane and between $1.09 : 1$ and $1.25 : 1$ perpendicular to the equatorial plane. This is in agreement with the results of Rasio & Shapiro (1994) and Zhuge et al (1994) who found this structure for models with stiff equations of state and adiabatic exponents $\Gamma \gtrsim 2.3$.

The peak luminosity of gravitational waves occurs shortly after the dynamical instability of the coalescing binary has set in and the neutron stars start do merge. A maximum luminosity in excess of $10^{55}$ erg/s is reached for about 1 ms. Although the energy emission rate drops after this short but powerful outburst because the merged object has a much smaller asymmetry than the pre-merging configuration, a significant contribution to the gravitational wave signal (about 50% of the total emitted energy), in particular at high frequencies $\gtrsim 1$ KHz, is radiated at later times. The total energy emitted in gravitational waves varies between 1.7%–2.2% of $M_\odot c^2$, which is considerably less than the values of 7–9% of $M_\odot c^2$ reported by Nakamura & Oohara (1991). The energy release in gravitational waves depends strongly on the size and compactness of the neutron stars since the gravitational wave luminosity increases like $1/a^5$ with the minimum separation of the neutron stars before they merge and before the binary system loses its highly non-spherical geometry. With a radius of about 9 km the polytropic neutron star models of Nakamura & Oohara (1991) are much smaller than those of our work (about 15 km) using the equation of state of Lattimer & Swesty (1991). Correspondingly, the gravitational wave amplitudes found in our simulations reach $3 \cdot 10^{-23}$ for a distance of 1 Gpc, while Shibata et al (1992), like Nakamura & Oohara (1991) having more compact neutron stars and using polytropic equations of state, obtained maximum wave amplitudes of $3.5$–$4 \cdot 10^{-23}$. Most of the gravitational wave energy is radiated in a window of frequencies between 1 KHz and 2 KHz which corresponds to the dynamical frequencies of the binary system and merged object.

The phase of the maximum emission of gravitational waves from the time when the neutron star surfaces begin to touch to the moment when their density maxima merge, is accompanied by a steep increase of the gravitational wave amplitude, followed by an abrupt drop and a subsequent extended period where the radiated gravitational waves exhibit a periodic modulation and oscillatory behavior with indications of the superposition of modes of different frequencies. This behavior depends on the triaxiality of the compact, merged object and is caused by the wobbling and ringing of the central, massive body after the dynamical coalescence. The details of the mass flow and therefore the structure of the gravitational waveforms are sensitive to the amount of angular momentum in the system and the corresponding degree of triaxial deformation and asymmetry. Positive correlations between the stiffness of the neutron star equation of state and the nonaxisymmetry of the final, merged object on the one side and the strength and amplitude of the post-merging oscillations of the gravitational wave emission on the other were pointed out by Rasio & Shapiro (1994) and Zhuge et al (1994). From our simulations, all performed with the same equation of state, we conclude that also the total angular momentum in the system affects the structure of the gravitational waveforms and determines the modulation of their amplitudes. In particular, the size of the gravitational wave amplitude, the peak luminosity of the gravitational waves, the strength of the post-merging emission, and the form of the gravitational wave energy spectrum $\frac{\mathrm{d}\mathcal{E}}{\mathrm{d}f}$, i.e., the energy emitted in gravitational waves at different frequencies, clearly depend on the rotational state of the merging system and of the coalesced object. Clear trends with the total angular momentum of the system are visible for some quantities, e.g., an increase of the maximum gravitational wave amplitude, an increase of the energy emitted in the 0.8–1.5 KHz band, and a decrease of the narrow spectral peak at a frequency of about 2 KHz. The temporal modulation of the waveforms and the power spectrum of the wave amplitude, however, seem to vary in a complex and non-trivial way and reflect details of the mass motions in the merger.

Zhuge et al (1994) have carefully analysed their models for the sensitivity of the gravitational wave energy spectra to the neutron star radii and the adiabatic exponent of the polytropic equation of state. They investigated the possibility to deduce information from the heights and frequencies of the spectral peaks on these neutron star properties. Our energy spectra are in gross agreement with the principal features seen in their run 2 which simulates the coalescence of two neutron stars with 15 km radius. In particular, we recover the gradual drop of the spectrum below the point-mass inspiral value near the frequency at which the dynamical instability sets in, and the existence of a primary peak associated with the main coalescence and of a secondary peak at higher frequency as a consequence of oscillations that occur during and after the merging. Also, the characteristic frequencies at which these peaks appear in our simulations match well with those of the Zhuge et al (1994) computation. The detailed structure of the energy spectrum, however, depends very sensitively on the total angular momentum of the system and therefore on the spins of the merging neutron stars. Moreover, the sharpness of the spectral structures seen by Zhuge et al (1994) is barely reproduced by our results. This may be due to the inclusion of effects from gravitational radiation (back)reaction in our models, but can also be a consequence of the bulk viscosity of the nuclear medium which is caused by shifts of the $\beta$-equilibrium and the corresponding production of neutrinos during compression and expansion phases. Note that since the disk is semi-transparent to neutrinos and the central, massive object is not completely opaque, neutrinos can escape and are not in perfect thermodynamical equilibrium with the stellar gas. Neutrino production and emission therefore give rise to dissipation. Both neutrino processes and gravitational radiation backreaction may be responsible for damping mass motions and smoothing fine structures. Finally, we point out that the shape of the emitted gravitational wave energy spectrum is strongly influenced by the duration of the intermediate phase of the stability of the merged object before its likely collapse into a black hole occurs. We find a drastic variation of the energy spectra with time and the spectra at about 3–5 ms after the start of our simulations (the expected time of the instability against gravitational collapse) are largely different from those obtained from integration until the end of our computations ($t \approx 10$ ms).

Inspiral and final merger of binary neutron stars (each of the neutron stars could also be replaced by a black hole) is one of the most promising types of sources which might be detected by the gravitational wave interferometers that are currently planned or under construction (see, e.g., Thorne, 1992). The maximum sensitivity of the LIGO experiment lies well below 1 KHz (namely around 100 Hz), which corresponds to an orbital distance of the neutron star binary of approximately 35 km. Therefore most of the interesting signal is emitted during the inspiral phase when the orbit of the neutron stars is still larger than 35 km and can be described by using the point-mass approximation with a correction term to account for the finite extent of the neutron stars. When the binary separation is comparable to the neutron star radius and the orbit gets close to the point of dynamical instability, hydrodynamic effects become important and the coalescence proceeds much different from the point-mass evolution. During the final stage of the merging the inspiraling binary neutron stars produce gravitational waves at dominant frequencies between 1 and 2 KHz. Although these frequencies lie outside the window of the largest sensitivity of the LIGO interferometers, the gravitational wave signals should nevertheless be detectable out to distances of several hundred Mpc. The emitted gravitational wave forms and spectra emitted during the merging of the binary are not only sensitive to the neutron star properties and the character of the equation of state of neutron star matter. The hydrodynamics and mass motions depend on the angular momentum and, to a somewhat lesser extent, also on microphysical processes in the merging stars. To extract valuable information about the merging event from the measured signals therefore requires detailed numerical models. In particular, steps away from the still basically Newtonian hydrodynamic models towards (fully) relativistic simulations need to be taken.

In a forthcoming paper we shall present a detailed evaluation of our models for the neutrino emission and will study the implications for $\gamma$-ray burster models. Neutrino-antineutrino annihilation in the surroundings of the merging neutron stars during the last stages of the coalescence has been suggested as a possible mechanism to create relativistic $e^+e^-$ fireballs which could be an efficient source of $\gamma$-rays (see, e.g., Cavallo & Rees, 1978; Paczyński, 1990; Narayan et al, 1992; Mészáros & Rees, 1992; Mochkovitch et al, 1993). In future we intend to extend our simulations to cover coalescence of neutron stars with different masses, collisions of neutron stars that are initially not in orbit around each other, and also the merger of a neutron star with a black hole where we shall attempt to include the main effects of the radius of the last stable orbit.

Movies in mpeg format of the dynamical evolution of all models are available in the WWW at `http://www.mpa-garching.mpg.de/~mor/nsgrb.html`

*Acknowledgements.* The calculations were performed at the Rechenzentrum Garching on a Cray-YMP 4/64 and a Cray-EL98 4/256. It is a pleasure to thank Wolfgang Keil for transforming Lattimer & Swesty's FORTRAN equation of state into a usable table and M.R. would like to thank Sabine Schindler for her patience in our office. H.-Th. J. would like to thank W. Hillebrandt for inspiring discussions. H.-Th. J. was supported in part by the National Science Foundation under grant NSF AST 92-17969, by the National Aeronautics and Space Administration under grant NASA NAG 5-2081, and by an Otto Hahn Postdoctoral Scholarship of the Max-Planck-Society. M.R. acknowledges support by the Deutsche Agentur für Raumfahrtangelegenheiten (DARA), Förderungsvorhaben des Bundes, FKZ: 50 OR 92095.

## A. Neutrino opacities

In this appendix the formulae are collected to compute neutrino opacities $\kappa = \lambda^{-1}$ (inverse mean free paths), optical depths, and diffusion timescales for all types of neutrinos and for the neutrino processes listed in Eq. (31)–Eq. (33).

The transport cross section (cross section for momentum transfer) for neutrino-nucleon scatterings is given by

$$\sigma_s(\nu_i N) = C_{s,N}\,\sigma_0 \left(\frac{\epsilon}{m_e c^2}\right)^2, \tag{A1}$$

where $\epsilon$ is the neutrino energy, $m_e$ is the electron rest mass, $c$ the speed of light, $\sigma_0 = 1.76 \times 10^{-44}\,\mathrm{cm}^2$, and $C_{s,n} = \left(1 + 5\alpha^2\right)/24$ for neutrino-neutron scattering and $C_{s,p} = \left[4(C_V - 1)^2 + 5\alpha^2\right]/24$ for neutrino-proton scattering with $C_V = \frac{1}{2} + 2\sin^2\theta_W$, $\sin^2\theta_W \approx 0.23$, and $\alpha \approx 1.25$.

We assume that the neutrino spectra can be represented by Fermi-Dirac distributions for the temperature $T$ (taken equal to the gas temperature) and neutrino degeneracy $\eta_{\nu_i} = \mu_{\nu_i}/T$ ($\mu_{\nu_i}$ being the neutrino chemical potential). $\eta_{\nu_i}$ is chosen to be

$$\eta_{\nu_x} = 0 \tag{A2}$$

for heavy-lepton neutrinos and

$$\eta_{\nu_e} = \eta_{\nu_e}^{\mathrm{ceq}} \cdot \left[1 - \exp(-\tau_{\nu_e,0})\right] + \eta_{\nu_e}^0 \cdot \exp(-\tau_{\nu_e,0}) \tag{A3}$$

for electron neutrinos and

$$\eta_{\bar\nu_e} = -\eta_{\nu_e}^{\mathrm{ceq}} \cdot \left[1 - \exp(-\tau_{\bar\nu_e,0})\right] + \eta_{\bar\nu_e}^0 \cdot \exp(-\tau_{\bar\nu_e,0}) \tag{A4}$$

for electron antineutrinos. $\eta_{\nu_e}^{\mathrm{ceq}}$ is the degeneracy parameter for $\nu_e$ at chemical equilibrium with the stellar medium,

$$\eta_{\nu_e}^{\mathrm{ceq}} = -\eta_{\bar\nu_e}^{\mathrm{ceq}} = \eta_e + \eta_p - \eta_n - Q/T, \tag{A5}$$

when $\eta_e$ is the degeneracy parameter of electrons (including electron rest mass contributions), $\eta_p$ and $\eta_n$ are the degeneracy parameters of $p$ and $n$, respectively (without rest masses), and $Q = 1.2935\,\mathrm{MeV}$ is the rest-mass-energy difference between a neutron and a proton. To avoid divergent behavior at low densities we interpolate for $\nu_e$ and $\bar\nu_e$ between the chemical equilibrium values at high optical depths and fixed values $\eta_{\nu_e}^0$ and $\eta_{\bar\nu_e}^0$ in the limit of transparent matter. We use $\eta_{\nu_e}^0 = \eta_{\bar\nu_e}^0 = 0$. The interpolations between both limits are expressed in terms of the optical depths $\tau_{\nu_e,0}$ and $\tau_{\bar\nu_e,0}$ for $\nu_e$ and $\bar\nu_e$ number transport (see below). Since the computation of optical depths already requires the knowledge of neutrino opacities, a first iteration step is performed where the deviation of $\eta_{\nu_e}$ and $\eta_{\bar\nu_e}$ from their chemical equilibrium values is simply written as a function of decreasing density $\rho$.

With defined neutrino spectra the spectrally averaged scattering opacities are now computed as

$$\kappa_{s,j}(\nu_i N) = C_{s,N}\,\sigma_0\,\mathcal{A}\,\rho\,Y_{NN}\left(\frac{T}{m_e c^2}\right)^2 \cdot \frac{\mathcal{F}_{4+j}(\eta_{\nu_i})}{\mathcal{F}_{2+j}(\eta_{\nu_i})}. \tag{A6}$$

For $j = 0$ one obtains the opacities for neutrino-number transport, for $j = 1$ the opacities for neutrino-energy transport. $\mathcal{A}$ is Avogadro's constant, $T$ the gas temperature (measured in energy units) and $\mathcal{F}_k(\eta) \equiv \int_0^\infty \mathrm{d}x\, x^k / \left[1 + \exp(x - \eta)\right]$ are the Fermi integrals for relativistic particles. $Y_{NN}$ means the number fractions of free neutrons $n$ or protons $p$, respectively, with the double index "$NN$" indicating that Pauli blocking effects for degenerate nucleons are taken into account by following Bruenn (1985), who suggested the replacements

$$Y_N \longrightarrow Y_{NN} = \begin{cases} Y_N & \text{for nondegenerate } N\,; \\ Y_N\,/\tfrac{2}{3}\eta_N & \text{for degenerate } N\,. \end{cases} \tag{A7}$$

Simple interpolation of both limits yields the combined formula

$$Y_{NN} = \frac{Y_N}{1 + \frac{2}{3}\max(\eta_N, 0)}. \tag{A8}$$

For completely dissociated matter the nucleon fractions are $Y_n = 1 - Y_e$ and $Y_p = Y_e$.

The cross sections for $\nu_e$ absorption onto $n$ in hot neutron star matter is given by

$$\sigma_a(\nu_e n) = \frac{1 + 3\alpha^2}{4} \sigma_0 \left(\frac{\epsilon}{m_e c^2}\right)^2 [1 - f_{e^-}(\epsilon)] , \tag{A9}$$

and for $\bar{\nu}_e$ absorption onto $p$ it is

$$\sigma_a(\bar{\nu}_e p) = \frac{1 + 3\alpha^2}{4} \sigma_0 \left(\frac{\epsilon}{m_e c^2}\right)^2 [1 - f_{e^+}(\epsilon)] \tag{A10}$$

(Tubbs & Schramm, 1975; Bruenn, 1985), where the factors in brackets are supposed to account for Pauli blocking effects in the electron and positron phase spaces, respectively. Performing the spectral averaging one finds the opacities for absorption processes as

$$\kappa_{a,j}(\nu_e n) = \frac{1 + 3\alpha^2}{4} \sigma_0 \mathcal{A} \rho Y_{np} \left(\frac{T}{m_e c^2}\right)^2 \cdot \\ \frac{\mathcal{F}_{4+j}(\eta_{\nu_e})}{\mathcal{F}_{2+j}(\eta_{\nu_e})} \cdot \langle 1 - f_{e^-}(\epsilon) \rangle \tag{A11}$$

and

$$\kappa_{a,j}(\bar{\nu}_e p) = \frac{1 + 3\alpha^2}{4} \sigma_0 \mathcal{A} \rho Y_{pn} \left(\frac{T}{m_e c^2}\right)^2 \cdot \\ \frac{\mathcal{F}_{4+j}(\eta_{\bar{\nu}_e})}{\mathcal{F}_{2+j}(\eta_{\bar{\nu}_e})} \cdot \langle 1 - f_{e^+}(\epsilon) \rangle . \tag{A12}$$

Again, the doubly indexed number fractions $Y_{np}$ and $Y_{pn}$ indicate the inclusion of Fermion blocking effects in the nucleon phase spaces (Bruenn 1985) and are defined as

$$Y_{np} = \frac{2Y_e - 1}{\exp(\eta_p - \eta_n) - 1} \tag{A13}$$

and

$$Y_{pn} = \exp(\eta_p - \eta_n) \cdot Y_{np} \tag{A14}$$

when $Y_n = 1 - Y_e$ and $Y_p = Y_e$ are used. For the phase space blocking of the leptons we employ the approximate expressions

$$\langle 1 - f_{e^-}(\epsilon) \rangle \cong \left\{ 1 + \exp\left[-\left(\frac{\mathcal{F}_5(\eta_{\nu_e})}{\mathcal{F}_4(\eta_{\nu_e})} - \eta_e\right)\right] \right\}^{-1} \tag{A15}$$

and

$$\langle 1 - f_{e^+}(\epsilon) \rangle \cong \left\{ 1 + \exp\left[-\left(\frac{\mathcal{F}_5(\eta_{\bar{\nu}_e})}{\mathcal{F}_4(\eta_{\bar{\nu}_e})} + \eta_e\right)\right] \right\}^{-1} . \tag{A16}$$

From the spectral averages of scattering and absorption opacities total transport opacities are computed for $\nu_e$,

$$\kappa_{t,j}(\nu_e) = \kappa_{s,j}(\nu_e n) + \kappa_{s,j}(\nu_e p) + \kappa_{a,j}(\nu_e n) , \tag{A17}$$

for $\bar{\nu}_e$,

$$\kappa_{t,j}(\bar{\nu}_e) = \kappa_{s,j}(\bar{\nu}_e n) + \kappa_{s,j}(\bar{\nu}_e p) + \kappa_{a,j}(\bar{\nu}_e p) , \tag{A18}$$

and for the heavy-lepton neutrinos $\nu_x$,

$$\kappa_{t,j}(\nu_x) = \kappa_{s,j}(\nu_x n) + \kappa_{s,j}(\nu_x p) . \tag{A19}$$

These allow us to compute optical depths along paths $[s_1, s_2]$ according to

$$\tau_{\nu_i,j}([s_1, s_2]) = \int_{s_1}^{s_2} ds \, \kappa_{t,j}(\nu_i) \tag{A20}$$

which then yields the diffusion timescale along the path as

$$t_{\nu_i,j}^{\text{diff}}([s_1, s_2]) \approx \frac{3(s_2 - s_1)}{c} \tau_{\nu_i,j}([s_1, s_2]) . \tag{A21}$$

The numerical factor in Eq. (A21) is of order unity but its exact value cannot be fixed from theoretical arguments. It must therefore be adjusted in course of a calibration of the neutrino leakage scheme with more precise transport methods. In three-dimensional situations the optical depths and diffusion timescales for all cells of the grid into all directions ($\pm x$, $\pm y$, $\pm z$) are evaluated. The diffusion timescale attributed to a particular cell is then chosen as the minimum of these diffusion times.

## B. Neutrino emission rates

Given the neutrino diffusion timescales the effective neutrino emissivities of all cells of our computational grid can be determined for the processes of Eq. (27)–Eq. (30).

The emission rate of $\nu_e$ (cm$^{-3}$s$^{-1}$) by the $\beta$-process of Eq. (27) in hot neutron star matter is given by

$$R_\beta(\nu_e) = \frac{1 + 3\alpha^2}{8} \frac{\sigma_0 c}{(m_e c^2)^2} \mathcal{A} \rho Y_{pn} \cdot \tilde{\varepsilon}_{e^-} \cdot \langle 1 - f_{\nu_e}(\epsilon) \rangle_\beta \tag{B1}$$

and the corresponding emission rate of $\bar{\nu}_e$ in process Eq. (28) is

$$R_\beta(\bar{\nu}_e) = \frac{1 + 3\alpha^2}{8} \frac{\sigma_0 c}{(m_e c^2)^2} \mathcal{A} \rho Y_{np} \cdot \tilde{\varepsilon}_{e^+} \cdot \langle 1 - f_{\bar{\nu}_e}(\epsilon) \rangle_\beta \tag{B2}$$

(Tubbs & Schramm, 1975; Bruenn, 1985). $Y_{np}$ and $Y_{pn}$ are defined in Eq. (A13) and Eq. (A14), respectively, and the blocking in the neutrino phase spaces is approximately taken into account by

$$\langle 1 - f_{\nu_e}(\epsilon) \rangle_\beta \cong \left\{ 1 + \exp\left[-\left(\frac{\mathcal{F}_5(\eta_e)}{\mathcal{F}_4(\eta_e)} - \eta_{\nu_e}\right)\right] \right\}^{-1} \tag{B3}$$

and

$$\langle 1 - f_{\bar{\nu}_e}(\epsilon) \rangle_\beta \cong \left\{ 1 + \exp\left[-\left(\frac{\mathcal{F}_5(-\eta_e)}{\mathcal{F}_4(-\eta_e)} - \eta_{\bar{\nu}_e}\right)\right] \right\}^{-1} , \tag{B4}$$

where the Fermi functions are evaluated with the average energies of captured electrons or positrons (approximately equal to the energies of produced $\nu_e$ or $\bar{\nu}_e$). In Eq. (B1) and Eq. (B2) and in the rates given further below energy moments of the electron and positron distributions occur, being defined as

$$\varepsilon_{e^\mp} \equiv \frac{8\pi}{(hc)^3} \cdot T^4 \cdot \mathcal{F}_3(\pm\eta_e) \,, \tag{B5}$$

$$\tilde{\varepsilon}_{e^\mp} \equiv \frac{8\pi}{(hc)^3} \cdot T^5 \cdot \mathcal{F}_4(\pm\eta_e) \,, \tag{B6}$$

and

$$\varepsilon^*_{e^\mp} \equiv \frac{8\pi}{(hc)^3} \cdot T^6 \cdot \mathcal{F}_5(\pm\eta_e) \,. \tag{B7}$$

The emission of $\nu_e$ or $\bar{\nu}_e$ by electron-positron pair annihilation (Eq. (29)) is given by

$$R_{ee}(\nu_e, \bar{\nu}_e) = \frac{(C_1+C_2)_{\nu_e \bar{\nu}_e}}{36} \frac{\sigma_0 c}{(m_e c^2)^2} \cdot \varepsilon_{e^-} \, \varepsilon_{e^+} \cdot$$
$$\langle 1 - f_{\nu_e}(\epsilon)\rangle_{ee} \, \langle 1 - f_{\bar{\nu}_e}(\epsilon)\rangle_{ee} \tag{B8}$$

(Cooperstein et al, 1986, 1987) with the $\nu_e$ and $\bar{\nu}_e$ phase space blocking factors being evaluated with the average energies of neutrinos produced by $e^+ e^-$-annihilation:

$$\langle 1 - f_{\nu_i}(\epsilon)\rangle_{ee} \cong$$
$$\left\{ 1 + \exp\left[ -\left( \frac{1}{2} \frac{\mathcal{F}_4(\eta_e)}{\mathcal{F}_3(\eta_e)} + \frac{1}{2} \frac{\mathcal{F}_4(-\eta_e)}{\mathcal{F}_3(-\eta_e)} - \eta_{\nu_i} \right) \right] \right\}^{-1} . \tag{B9}$$

The weak interaction constants are $(C_1+C_2)_{\nu_e \bar{\nu}_e} = (C_V - C_A)^2 + (C_V + C_A)^2$ with $C_A = \frac{1}{2}$. For the production of $\nu_\mu$, $\nu_\tau$, $\bar{\nu}_\mu$, and $\bar{\nu}_\tau$ the corresponding rate is

$$R_{ee}(\nu_x) = \frac{(C_1+C_2)_{\nu_x \bar{\nu}_x}}{9} \frac{\sigma_0 c}{(m_e c^2)^2} \cdot \varepsilon_{e^-} \, \varepsilon_{e^+} \cdot$$
$$\left( \langle 1 - f_{\nu_x}(\epsilon)\rangle_{ee} \right)^2 \tag{B10}$$

with $(C_1+C_2)_{\nu_x \bar{\nu}_x} = (C_V - C_A)^2 + (C_V + C_A - 2)^2$.

The rate of creation of $\nu_e$ or $\bar{\nu}_e$ by the decay of transversal plasmons (Eq. (30)) (longitudinal plasmons yield a negligible contribution, see Schinder et al. 1987) can be written with sufficient accuracy as

$$R_\gamma(\nu_e, \bar{\nu}_e) \approx \frac{\pi^3}{3\alpha^*} C_V^2 \frac{\sigma_0 c}{(m_e c^2)^2} \cdot \frac{T^8}{(hc)^6} \cdot$$
$$\gamma^6 \, e^{-\gamma} \, (1+\gamma) \cdot \langle 1 - f_{\nu_e}(\epsilon)\rangle_\gamma \, \langle 1 - f_{\bar{\nu}_e}(\epsilon)\rangle_\gamma \tag{B11}$$

and the corresponding rate for producing $\nu_\mu$, $\bar{\nu}_\mu$, $\nu_\tau$, and $\bar{\nu}_\tau$ is

$$R_\gamma(\nu_x) \approx \frac{4\pi^3}{3\alpha^*} (C_V - 1)^2 \frac{\sigma_0 c}{(m_e c^2)^2} \cdot \frac{T^8}{(hc)^6} \cdot$$
$$\gamma^6 \, e^{-\gamma} \, (1+\gamma) \cdot \left( \langle 1 - f_{\nu_x}(\epsilon)\rangle_\gamma \right)^2 \, . \tag{B12}$$

$\alpha^*$ is the fine-structure constant, $\alpha^* = 1/137.036$, and $\gamma \approx \gamma_0 \sqrt{\frac{1}{3} (\pi^2 + 3\eta_e^2)}$ with $\gamma_0$ being related to the plasma frequency by $\gamma_0 = \hbar\Omega_0/m_e c^2 = 2\sqrt{\alpha^*/(3\pi)} = 5.565 \cdot 10^{-2}$. The average blocking factor for $\nu_i$ production can be estimated by evaluating

the Fermi function $f_{\nu_i}(\epsilon)$ with the average energy of neutrinos created by the plasmon decays:

$$\langle 1 - f_{\nu_i}(\epsilon)\rangle_\gamma \cong \left\{ 1 + \exp\left[ -\left( 1 + \frac{1}{2} \frac{\gamma^2}{1+\gamma} - \eta_{\nu_i} \right) \right] \right\}^{-1} . \tag{B13}$$

The neutrino-energy emission rates by the $\beta$-processes, $e^+ e^-$-pair annihilation, and plasmon decay are

$$Q_\beta(\nu_e) = R_\beta(\nu_e) \cdot \frac{\varepsilon^*_{e^-}}{\tilde{\varepsilon}_{e^-}} \,, \tag{B14}$$

$$Q_\beta(\bar{\nu}_e) = R_\beta(\bar{\nu}_e) \cdot \frac{\varepsilon^*_{e^+}}{\tilde{\varepsilon}_{e^+}} \,, \tag{B15}$$

$$Q_{ee}(\nu_i) = R_{ee}(\nu_i) \cdot \frac{1}{2} \frac{\tilde{\varepsilon}_{e^-}\varepsilon_{e^+} + \varepsilon_{e^-}\tilde{\varepsilon}_{e^+}}{\varepsilon_{e^-}\,\varepsilon_{e^+}} \,, \tag{B16}$$

and

$$Q_\gamma(\nu_i) = R_\gamma(\nu_i) \cdot \frac{1}{2} \, T \left( 2 + \frac{\gamma^2}{1+\gamma} \right) \tag{B17}$$

for neutrino species $\nu_i$.

Making use of the neutrino number densities,

$$n_{\nu_i} \equiv g_{\nu_i} \frac{4\pi}{(hc)^3} \, T^3 \cdot \mathcal{F}_2(\eta_{\nu_i}) \,, \tag{B18}$$

and energy densities,

$$\varepsilon_{\nu_i} \equiv g_{\nu_i} \frac{4\pi}{(hc)^3} \, T^4 \cdot \mathcal{F}_3(\eta_{\nu_i}) \,, \tag{B19}$$

where the numerical multiplicity factor is $g_{\nu_e} = g_{\bar{\nu}_e} = 1$ and $g_{\nu_x} = 4$, we can define neutrino-number emission timescales $t^{\text{loss}}_{\nu_i,0}$ as

$$\left( t^{\text{loss}}_{\nu_i,0} \right)^{-1} \equiv \frac{1}{n_{\nu_i}} \left[ R_\beta(\nu_i) + R_{ee}(\nu_i) + R_\gamma(\nu_i) \right] \equiv \frac{R(\nu_i)}{n_{\nu_i}} \tag{B20}$$

and neutrino-energy emission timescales $t^{\text{loss}}_{\nu_i,1}$ as

$$\left( t^{\text{loss}}_{\nu_i,1} \right)^{-1} \equiv \frac{1}{\varepsilon_{\nu_i}} \left[ Q_\beta(\nu_i) + Q_{ee}(\nu_i) + Q_\gamma(\nu_i) \right] \equiv \frac{Q(\nu_i)}{\varepsilon_{\nu_i}} \tag{B21}$$

(of course, the $\beta$-processes only apply for $\nu_e$ and $\bar{\nu}_e$). *Effective emission rates* of neutrino number and energy can now be defined as functions of the relative size of the *shortest* diffusion timescale $t^{\text{diff}}_{\nu_i,j}$ of a grid cell and the neutrino emission timescale $t^{\text{loss}}_{\nu_i,j}$ of this cell:

$$R^{\text{eff}}(\nu_i) \equiv \frac{R(\nu_i)}{1 + t^{\text{diff}}_{\nu_i,0} \cdot \left( t^{\text{loss}}_{\nu_i,0} \right)^{-1}} \,, \tag{B22}$$

$$Q^{\text{eff}}(\nu_i) \equiv \frac{Q(\nu_i)}{1 + t^{\text{diff}}_{\nu_i,1} \cdot \left( t^{\text{loss}}_{\nu_i,1} \right)^{-1}} \,. \tag{B23}$$

In the (nearly) transparent, low-density regime the optical depth to neutrinos vanishes and the diffusion timescale is short compared to the timescale of direct neutrino loss. Therefore $R^{\mathrm{eff}}(\nu_i) \to R(\nu_i)$ and $Q^{\mathrm{eff}}(\nu_i) \to Q(\nu_i)$. On the other hand, in the opaque region the diffusion timescale becomes very long (and due to the increase of the rates with density and temperature the direct loss timescale becomes extremely short, too). In this case $R^{\mathrm{eff}}(\nu_i) \to n_{\nu_i}/t^{\mathrm{diff}}_{\nu_i,0}$ and $Q^{\mathrm{eff}}(\nu_i) \to \varepsilon_{\nu_i}/t^{\mathrm{diff}}_{\nu_i,1}$, i.e. the loss of neutrinos is then determined by the much slower diffusion process that depletes the equilibrium neutrino distributions only on a long time.

Finally, we define the total electron-lepton number loss rate of the stellar gas as

$$R^-_{Y_e} \equiv -R^{\mathrm{eff}}(\nu_e) + R^{\mathrm{eff}}(\bar{\nu}_e) \tag{B24}$$

and the total energy loss rate as

$$Q^-_\varepsilon \equiv -\left[Q^{\mathrm{eff}}(\nu_e) + Q^{\mathrm{eff}}(\bar{\nu}_e) + Q^{\mathrm{eff}}(\nu_x)\right] . \tag{B25}$$

Eq. (B24) and Eq. (B25) are the source terms to be used in the continuity equation for electron-lepton number (Eq. 14: $S_{\mathrm{L}} \equiv R^-_{Y_e}$) and in the gas-energy equation (Eq. 3: $S_{\mathrm{E}} \equiv Q^-_\varepsilon$), respectively.

The average energy of neutrinos of species $\nu_i$ emitted from a grid cell is given by

$$\langle \epsilon \rangle_{\nu_i} = \frac{Q^{\mathrm{eff}}(\nu_i)}{R^{\mathrm{eff}}(\nu_i)} = \frac{Q(\nu_i)}{R(\nu_i)} \cdot \frac{1 + t^{\mathrm{diff}}_{\nu_i,0} \cdot \left(t^{\mathrm{loss}}_{\nu_i,0}\right)^{-1}}{1 + t^{\mathrm{diff}}_{\nu_i,1} \cdot \left(t^{\mathrm{loss}}_{\nu_i,1}\right)^{-1}} \tag{B26}$$

and the average energy of neutrinos $\nu_i$ emitted from the star is determined as the ratio of the rates $Q^{\mathrm{eff}}(\nu_i)$ and $R^{\mathrm{eff}}(\nu_i)$ summed individually over the whole grid.